# Human-centered AI with focus on Human-robot interaction

Alireza Mortezapour*

Department of Computer Science, University of Salerno, 84084Fisciano, Italy

Email: amortezapoursoufiani@unisa.it / amortezapour258@gmail.com

ORCID: 0000-0001-6356-2244

Giuliana Vitiello

Department of Computer Science, University of Salerno, 84084Fisciano, Italy

Email: gvitiello@unisa.it

ORCID: 0000-0001-7130-996X

**Abstract:**

Modern social robots can be considered the descendants of steam engines from the First Industrial Revolution (IR 1.0) and industrial robotic arms from the Third Industrial Revolution (IR 3.0). As some time has passed since the introduction of these robots during the Fourth Industrial Revolution (IR 4.0), challenges and issues in their interaction with humans have emerged, leading researchers to conclude that, like any other AI-based technology, these robots must also be human-centered to meet the needs of their users. This chapter aims to introduce humans and their needs in interactions with robots, ranging from short-term, one-on-one interactions (micro-level) to long-term, macro-level needs at the societal scale. Building upon the principles of human-centered AI, this chapter presents, for the first time, a new framework of human needs called the **Dual Pyramid**. This framework encompasses a comprehensive list of human needs in robot interactions, from the most fundamental—robot effectiveness—to macro-level requirements, such as the collaboration with robots in achieving the United Nations' 17 Sustainable Development Goals.

**Keywords:** Human-Robot interaction; Human-centered AI; Human-AI symbiosis; Human needs pyramid; Human; Industry 5.0, Human-centered design; Human-AI interaction; Human-centered robot; Ergonomic Robot.

# 1. Introduction

## 1.1. Along the Path of Industrial Revolutions: Robots as the Offspring of Machines

When studying human-robot interaction, it is more appropriate to first become acquainted with the concept of the Industrial Revolution. In brief, the Industrial Revolution refers to a period in history during which, in response to population growth and the need for increased production, as well as the pursuit of higher productivity and profit, methods of production shifted from manual approaches to the use of machines and assembly lines in factories. During this period, which is considered to have begun in the late 18th century in England, production machines were at the heart of the Industrial Revolution. They were introduced to perform tasks previously done by humans, more quickly, more precisely, without fatigue, and in greater volumes.

In the First Industrial Revolution, these machines were powered by steam, and in the Second Industrial Revolution, by electricity. The Third Industrial Revolution emerged in the mid-20th century with the advent of computers and information technology, while the Fourth Industrial Revolution has unfolded with further advancements in computing and the emergence of artificial intelligence. In summary, while robots are indeed a product of the Fourth Industrial Revolution, it is more accurate to consider them as the offspring of the machines that supported workers during the First Industrial Revolution (Cheng et al., 2021).

In the past few decades, following the Third Industrial Revolution and with the advent of computers and various branches of engineering sciences, the concept of "robot" emerged as machines capable of intelligent, goal-oriented interactions with both humans and their environments. In other words, the term robot replaced the earlier notion of simple machines that initially operated using steam and later electricity. During this period, the first generation of industrial robots, or robotic arms capable of performing repetitive tasks with high precision, was introduced. Unsurprisingly, their application was confined to factories and mass production lines. The inception of this generation is attributed to a General Motors factory in New Jersey, USA, around 1961 (Babamiri et al., 2024a).

However, with scientific advancements, robots were no longer limited to factories. Today, with their enhanced capabilities for intelligent interaction with humans, they have entered society. The term "social robot" was first introduced in 1978 in the Interface Age magazine (Bartneck et al., 2024). Today, social robots directly interact with humans in hospitals, hotels, educational centers, recreational and shopping facilities, and even some homes. This marks a turning point where robots have expanded not merely to increase speed and productivity but to enhance quality of life, assist humans, and empower them in various settings (Cavallaro et al., 2024).

Naturally, their appearance has transformed from sometimes very large mechanical arms into human-centered social robots, sometimes just a few dozen centimeters tall. Additionally, the nature of their interaction with humans has undergone fundamental changes.

## 1.2. Bidirectional nature of Human-robot interaction

As mentioned earlier, the nature of human-robot interaction has significantly evolved from early industrial robots to today's social robots. In the past, the interaction was hierarchical: humans acted as commanders, and robots served as executors or operators. Today, thanks to remarkable technological advancements, robots are no longer mere executors. They not only execute human commands in various interactions but also assess and analyze their environments, learn from human behaviors, and, enhanced by integrated artificial intelligence, make independent decisions. They can present these decisions in a two-way interaction, inform humans, and even wait for their feedback to adjust their behavior or initiate a new interaction. It can be argued that, due to recent advancements, robots have transitioned from being merely tools to becoming collaborators. As highlighted, this interaction is no longer unidirectional. Interactions with earlier generations of robots were primarily limited to computer science and robotics specialists or programmers. However, with the introduction of generative AI models and various language and visual models, interacting with robots has become easier for non-expert individuals. Even through simple conversations, this two-way communication can now be established.

In summary, robots are no longer merely machines designed to perform tasks. Instead, they have become interactive agents that enhance human quality of life through mutual relationships. This transformation—from relatively simple tools to intelligent and interactive beings—places the concept of two-way human-robot communication at its core. This shift entails a transition from command-execution relationships to a more complex and dynamic level, where emotions and feelings are sometimes intertwined (Maniscalco et al., 2022).

Now that we understand the nature of changes in human-robot interaction in recent years, it is worthwhile to review the aspects from various scientific fields that have facilitated these changes or been influenced by them. The technological dimensions enabling this interaction can be summarized into several categories. The most significant among them is the ability of robots to analyze human emotions and behaviors. This capability, which forms the foundation of two-way interaction, has been made possible by advancements in technologies such as natural language processing, computer vision, and deep learning. The next step after initiating two-way interaction with humans is continuous learning through long-term interactions. In other words, robots must effectively update their information based on new needs and users' behavioral patterns during their interactions with humans and the environment. The overall framework of this interaction has also been enabled by advancements in and integration of multi-sensory technologies, including vision, sound, and even touch. This means that today's robots can utilize a combination of senses in their interactions (André, 2023).

Moving beyond the technologies that have enabled and facilitated two-way interactions, we must address the social and behavioral impacts of these interactions on communities. The presence of robots with two-way interaction capabilities has played a significant role in improving the quality of life for elderly individuals. These individuals, categorized as those with special needs within societies, can now, through interactions with these robots, assume more effective roles in communities and feel a greater sense of usefulness with the assistance of robots. This list, of course, extends beyond the elderly to include children, patients, individuals with disabilities, and others (Cavallaro et al., 2024).

Thus, it can be stated that the presence of robots at the grassroots level of society, through their two-way interaction capabilities, has already changed or will continue to change many behavioral patterns. Even among other segments of society, concepts such as trust in technology and the acceptance of robots as collaborators are directly tied to this two-way interaction. To justify changes in social norms, it is sufficient to note that with the presence of robots and their ability to interact with humans, people can now experience a new sensation—being understood by a machine—a concept that did not exist until recently. This, in turn, can serve as a pathway to improving social skills, reducing psychological stress, and enhancing the quality of life (S. Kim et al., 2022; Winkle et al., 2022).

On the other hand, these two-way interactions also raise new concerns, such as the erosion of personal privacy and the emergence of new social patterns that might not have been ethically or socially acceptable before.

### 1.3. How HRI differs from other form of Human-computer interaction?

As we considered robots as descendants of earlier machines, human-robot interaction can also be viewed as an evolved model of human-computer interaction. In this sense, humans interacting with robots are essentially engaging with a technology, much like previous generations of human-computer interaction. This section aims to highlight the differences between human-robot interaction and human interaction with other technologies. While it is true that human-robot interaction shares many similarities with human-computer interaction, this new form of interaction encompasses dimensions that are rarely found in other interactions—or, as some researchers have claimed, are unique to human-robot interaction.

#### 1.3.1. Physical Embodiment of Robots

In human-computer interaction, humans typically interact through touchscreens or indirectly via a mouse and keyboard. The prevailing belief is that these means of interaction lack physical embodiment and serve merely as intermediaries. In contrast, interaction with robots involves engaging with a technology that possesses physical embodiment, which, in some cases, can move within the physical world and share human spaces. This physical embodiment gives robots the ability to perform certain physical tasks, such as assisting humans, picking up objects, or exploring hazardous environments in place of rescuers, enabling successful interactions. Furthermore, this embodiment ties directly to another dimension—active agency. Robots, by perceiving environmental elements and user needs, can independently make decisions and perform purposeful activities within the physical environment (Sasser et al., 2024).

#### 1.3.2. Social Dimension and Human-like Behaviors

It is challenging to argue that earlier generations of human-computer interaction exhibited a social dimension or interactions resembling human behavior. However, human-robot interaction, particularly with anthropomorphic robots, fundamentally differs. In these interactions, there is no longer a need to rely on programming commands or predefined user interfaces to communicate with technology. Instead, robots can interact with humans using more social and human-like ways, such as body language, gestures, and facial expressions. They can also analyze users' behavior and appearance to understand their actions and establish meaningful communication.

Currently, significant efforts are being made by researchers to enhance robots' abilities in emotional perception and the expression of emotions. Reviewing recent studies reveals that researchers are also exploring the integration of more complex social interactions, such as ethical decision-making in specific scenarios. These efforts aim not only to develop robots capable of understanding users' emotional states but also of empathizing with them and assisting them in making complex ethical judgments within society. This capability would elevate robots from mere tools to agents that can actively contribute to enhancing human decision-making and strengthening social bonds (Ottoni & Cerqueira, 2024).

#### 1.3.3. Capability for Long-Term Interaction

In traditional human-computer interaction, long-term interaction was generally either absent or, when present, lacked the ability to learn and adapt from past interactions to improve future interaction.

It is the unique capability of social robots to learn from past experiences, correct previous errors, and thereby build greater trust and improve the user experience over time. This adaptability to users' evolving needs was rarely observed in older interaction models. Through continuous learning, social robots not only refine their responses but also enhance their ability to meet user expectations, opening the way for deeper, more meaningful, and personalized long-term interactions. This capability significantly distinguishes human-robot interaction from traditional forms of interaction with technology (Irfan et al., 2021).

**1.4. Expanding the role of robots to various domains:**

In the previous section, we highlighted how the essence of human-robot interaction stems from the robots' abilities to learn, adapt, and understand humans, alongside shifts in social and psychological concepts such as technology acceptance and trust in machines. This bidirectional interaction has led to the significant presence of robots in society. Robots are no longer confined to industries; reports about their integration into hospitals, schools, and even homes are becoming increasingly common.

It hasn't been long since humanoid robots from companies like Boston Dynamics, Tesla, and Digit (developed by Agility Robotics) were introduced to assist individuals in their homes. As IEEE Spectrum suggests, it is time to set aside the myth of dangerous and lethal robots and start focusing on the helpful robots we've always dreamed of—those designed to assist us and improve our lives. Before delving into the various dimensions of robots' presence in society, it is essential to emphasize that the primary distinction between these robots and industrial robots lies in the social interaction established between them and humans. Social robots have undergone significant transformations both in terms of their social abilities and their appearance.

The introduction of the first social robot dates back to around 1997 at MIT. One of the university's doctoral students developed the robot Kismet as part of the PhD dissertation. Kismet featured a head equipped with eyes, lips, eyebrows, and a neck, enabling it to respond to objects or individuals within its field of vision. It can be considered one of the earliest robots with the capability to establish social interactions (Ferrell & Kemp, 1996).

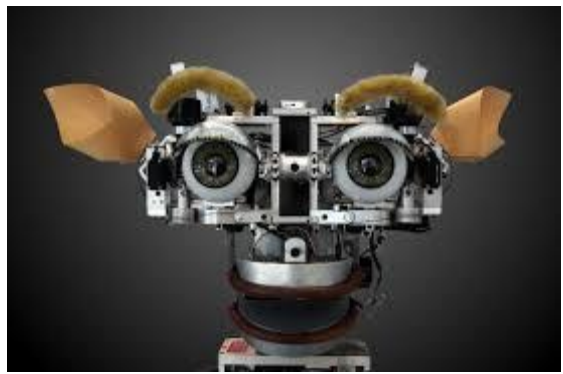

*Figure 1. Kismet robot from MIT University adopted from (Breazeal, 2000).*

With advancements in technology over the past 20 years, numerous other social robots have been introduced, showcasing stronger social interactions and the ability to function in diverse environments such as hospitals, hotels, and schools. Some of the most notable social robots developed to date include [NAO](), [Pepper](), [Keepon](), [Paro](), [Baxter](), [Aibo ERS-1000](), [Astro](), [InMoov](), and [Blossom](). This list continues to expand daily. In particular, the emergence of powerful AI models and related advancements in recent years has led to the development of highly human-like social robots. Examples of these include [Furhat]() from Sweden, [Ameca]() from the UK, [Surena]() from Iran, [Sophia]() from Hong Kong, [Navel]() from Germany, [Tesla Optimus]() from the USA, and [AI Buddy]() from France.

**1.4.1. Social robots in healthcare:**

When discussing the presence of social robots in healthcare environments, this typically refers to their use in hospitals, elderly care homes, mental health facilities, psychotherapy clinics, and rehabilitation or physiotherapy centers. These robots primarily serve roles such as therapy assistants, caregivers, medication distributors, and general health monitors (Cantone et al., 2023).

A review of the latest articles in this field highlights several successful applications of social robots, including: Medication distribution for patients, enhancing the meaningfulness of life for elderly individuals and boosting their sense of independence, assisting seniors with daily routines, providing physical rehabilitation services and supporting daily exercise routines, reducing reliance on human caregivers, facilitating communication with family members, delivering pet therapy-based treatments for patients, decreasing stress, anxiety, and pain for children undergoing medical procedures, offering psychological interventions (under the supervision of human psychologists), providing medical education to nursing and medical students, monitoring patients' health status and measuring physical data (González-González et al., 2021; Ragno et al., 2023).

These studies involve various social robots, amounting to a list of over 89 distinct models. The most notable include:

[NAO](), [Pepper](), [Op3](), [InMoov](), [KIRO](), [ARASH](), [RAPIRO](), [BRIAN](), [KISMET](), [NECORO](), [PARO](), [AIBO](), [HUGG Able](), [THERABOT](), [PLEO](), [SNUGGLEBOT](), [EDU'](), [MAYA](), [Temi](), [Moxi](), [Jibo](), [QTrobot](), [Kaspar](), [Buddy](), [ELLIQ](), and [Haptic Creature]().

It is evident that these robots differ in characteristics such as the level of embodiment, physical appearance, mobility capabilities and degrees of freedom, sensors, algorithms, and interaction patterns with humans. However, detailing these differences is beyond the scope of this chapter.

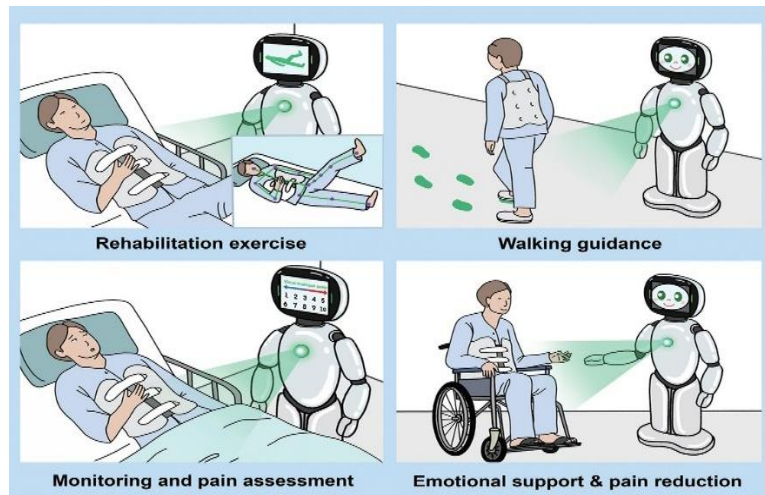

*Figure 2. Social robots in healthcare context. Adopted from (Han et al., 2024)*

**1.4.2. Social robots in educational context:**

For several years, the use of technology in educational assistance mainly revolved around simple tools such as PowerPoint presentations and video clips. However, thanks to the advancements brought about by the era of artificial intelligence, significantly more advanced technologies are now employed, including virtual reality headsets and social robots (Perillo et al., 2024).

Naturally, proponents of these technologies highlight benefits such as increased student motivation and engagement, enhanced social skills, personalized learning experiences, and reduced workload for teachers. On the other hand, numerous studies point to potential drawbacks, including threats to privacy, overdependence on technology, and the high costs associated with this mode of education (Johal, 2020).

Overall, the interaction between teachers and robots or students and robots has been examined in various studies through the use of over 50 different models of robots including: [NAO](), [Saya](), [RoboThespian](), [BIOLOID](), [Zenbo](), [Milo](), [Jibo](), [Reeti](), [XIAO](), [Spiderino](), [ONO](), [Aibo](), [Pepper](), [Zeno R25](), [Cozmo](), [Tega](), [Bender](), [SociBot Mini](), [Matilda](), [Robovie](), [Spykee](), and [Pleo]().

To assess the effectiveness and efficiency of social robots in educational environments, reviews of findings from recent meta-analyses and systematic reviews reveal key insights. One of the most recent meta-analyses, which synthesized results from 17 randomized studies on the impact of educational robots on learning outcomes, demonstrated a moderate effect size for the use of robots. This indicates that the presence of these robots showed significantly higher effectiveness compared to other technologies employed in education. The reported effect size in this study was 0.57, with a 95% confidence interval of 0.49 to 0.65, which was statistically significant. A closer examination of the results reveals that the presence of robots had a more pronounced influence on students' attitudes toward the subject matter than on final test scores or skill-based assessment indicators. Moreover, the effectiveness of robots was greater in higher educational levels, such as secondary and high school, compared to lower levels like primary and preschool (Wang et al., 2023).

In general, other systematic reviews also reported low to moderate positive effectiveness of robots. However, these studies emphasized several concerns, such as potential deterioration in human communication skills, privacy issues, and the risk of job displacement for teachers. Some researchers argue that due to the inherently interactive nature of social robots, they may be particularly suitable for teaching subjects like languages, where interaction plays a crucial role. Even under this assumption, a new meta-analysis found a moderate effect size for the use of robots in language education compared to the absence of any technological tools (d = 0.59, CI 95%: 0.41–0.76) (Lee & Lee, 2022).

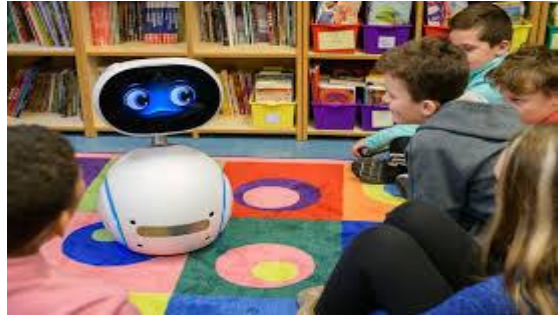

*Figure 3. Social robots in educational context. Adopted from (Schaffhauser, 2020)*

**1.4.3. Social robots in services (Hotels, Airports, Restaurants, Malls):**

With the advancement of navigation capabilities and obstacle detection to avoid collisions with moving individuals in dynamic environments, the presence of social robots in public spaces such as airports, museums, train stations, hotels, conference halls, and shopping centers has significantly increased. These robots are utilized to meet humans' social, security, and recreational needs. For instance, social robots in hotels are used to greet guests, provide initial guidance for reservations or room delivery, and even conduct satisfaction surveys. In museums, these robots interact with visitors, explain artistic and historical works, and sometimes engage in conversations in visitors' native languages. Similarly, in airports and train stations, they not only guide passengers but also assist with security tasks and essential checks. This category of robots is generally referred to as service robots. Among the many examples mentioned in earlier sections, notable focus is placed on robots like NAO and Pepper for such applications. While the benefits of these robots, such as enhancing user experience and improving customer satisfaction, are widely acknowledged, significant challenges like safety, data security, privacy, and ethical concerns are also central to researchers' discussions about their deployment (Gasteiger et al., 2021; Mubin et al., 2018).

**1.5. Why HRI needs to be optimized?**

With the efforts detailed in the previous sections, readers are expected to have gained an understanding of the general capabilities of social robots for integration into human environments. It was also demonstrated how these robots, with their immense diversity in models, appearances, and functionalities, have been developed to co-exist with humans. However, alongside the extensive efforts to design precise, high-tech, and quality robots, it is crucial to focus on the second aspect of interaction—the human dimension.

The belief is that no matter how well a robot is designed and manufactured, it must be optimized for interaction with humans (Babamiri et al., 2022). This is where concepts such as human acceptance of robots, user experience in working with robots, perceived safety, trust in robots, and many other social and cognitive aspects come into play. A robot is only deemed successful when its design, development, and implementation adequately address concepts related to human factors (as elaborated in the second part of this chapter) and when the robot is truly optimized for interaction with humans. In general, optimizing human-robot interaction encompasses various dimensions. In the third part of this chapter, we aim to delve deeper into one of the approaches to achieving this optimization: the human-centered perspective.

## 1.6. Objective and scope of the Chapter: Human-centered AI considerations in HRI

For over 70 years, researchers have strived to develop intelligent machines, powered by artificial intelligence, inspired by human intelligence. As mentioned earlier, nearly 55 years ago, the first generation of robots was introduced to the industry. Throughout these decades of developing and implementing intelligent machines—and especially in recent years with industrial and social robots—a wide range of benefits from these technologies has been introduced to society. In previous sections, key aspects of these benefits have been briefly reviewed.

However, alongside these advantages, artificial intelligence in general and robots, as its specialized offspring, have also brought various challenges. For example, concerns have been raised in the literature about workers fearing replacement by robots and consequent job losses (Babamiri et al., 2022). On a broader scale, the loss of human workforce skills has also been highlighted (Babamiri et al., 2024a). Other issues previously mentioned include ethical considerations, safeguarding users' private data when interacting with robots, high energy consumption, and the sustainability challenges associated with them (Caterino et al., 2024). Moreover, as large language models are increasingly integrated into robots, associated issues such as biases and algorithmic unfairness have added to the concerns in human-robot interaction.

Overall, it has been expected that AI-powered technologies—and in the context of this chapter, robots—could alleviate societal challenges such as poverty, unemployment, and inequality, in alignment with the 17 Sustainable Development Goals (SDGs) of the United Nations (Olaronke et al., 2022). While significant progress has been made, recent trends show that the development of these technologies has also introduced new challenges or exacerbated existing ones.

One of the critical solutions proposed to address these challenges is human-centered AI-based technologies. In recent years, researchers in the field of human-computer interaction have argued that a technology is human-centered if it is designed, developed, and utilized based on user experience principles. However, this understanding has matured in recent years. The concept of human-centeredness now extends beyond mere adherence to user experience design principles to encompass notions such as responsibility, trustworthiness, and sustainability (Sharma & Shrestha, 2024).

In essence, robots should be designed, developed, and used in ways that ensure not only efficient, effective, and satisfying human interaction (the baseline usability standards) but also promote broader concepts like accountability, social ethics, reducing inequality, well-being, and overall empowerment of users (He et al., 2021).

This chapter aims to focus on the concepts of human-centered AI in human-robot interaction. The pivotal contribution of this chapter is the introduction of a new framework for human-centering robots based on the human needs. This novel framework is constructed using two inverted pyramids stacked upon each other. It encompasses the minimum individual human needs for robot interaction at its base and extends to the maximum societal needs for interacting with large groups of robots to achieve the United Nations' Sustainable Development Goals.

## 2. Human Psychology principles and Sociology considerations in designing HRI:

When discussing a good user experience in interaction with robots, it is essential to understand that we are fundamentally talking about humans, their abilities, capabilities, and limitations. It is the human user who must report a positive user experience, which means that in the process of human-centering robot design, we need to gain a deeper understanding of humans, particularly the cognitive and sociological aspects relevant to them. This section of the chapter aims to provide an overview of humans and their characteristics that are essential in interactions with robots.

### 2.1. Human Psychology:

When studying the foundational concepts of cognition in humans for designing interactions, it is often sufficient to start by understanding that various states of cognition can be envisioned for a human including thinking, remembering, reading, learning, imagining, decision-making, seeing, writing, speaking, and more. Naturally, this list is not exhaustive and can be expanded to include other cognitive activities.

Researchers have proposed various classifications to categorize this wide range of cognitive concepts. For instance, Donald Norman, often referred to as the father of user experience, introduced a dual-mode classification of experiential cognition and reflective cognition (Norman, 1993). Similarly, Daniel Kahneman, a Nobel Prize laureate in Economics and a prominent psychologist, proposed the Dual-System Theory, dividing cognitive processes into fast thinking and slow thinking (Kahneman, 2011). From another perspective, Yvonne Rogers offered a three-mode categorization: individual cognition, social cognition, and distributed cognition (Rogers & Ellis, 1994). A related four-mode classification has also been introduced, encompassing distributed cognition, situated cognition, extended cognition, and embodied cognition (Risku & Rogl, 2020).

It is evident that these classifications are developed with different goals and approaches in mind. However, they are presented here to provide readers of this chapter with a better understanding of how cognitive processes can be categorized. Furthermore, some of these concepts will be utilized in subsequent sections to clarify human cognitive patterns in interaction with technology.

Alongside all these classifications, human cognitive processes can also be divided into several specific subprocesses, including attention, perception, memory, learning, and problem-solving/decision-making (Cross & Ramsey, 2021; Sarter & Sarter, 2003). Since most of these subprocesses often coexist during the cognitive activities and actions previously mentioned, we can expect them to be interwoven in human interaction with a robot. Assuming such an interaction is considered a single or interconnected set of cognitive actions, these subprocesses occur in an interconnected manner, enabling us to experience interaction with the robot.

**Attention** involves selecting among options available in the environment, enabling us to focus on sensory inputs and information essential to our needs amidst an overwhelming abundance of data. This ability protects us from being inundated with irrelevant information, which could otherwise lead to fatigue and frustration.

In human-robot interaction, attention encompasses the ability to identify, filter, and separate essential information from irrelevant details, focusing only on what is necessary for an optimal interaction. From this perspective, visually comprehensible and clear design on the part of robots can significantly enhance selective attention. Beyond selective attention, other types, such as divided attention, also play a critical role, especially in interactions involving groups of robots or a combination of humans and robots simultaneously. In such scenarios, users cannot disregard any information as irrelevant and must allocate their attention resources across multiple sources.

If robots present excessive amounts of information, overly complex feedback, or unclear signals, this can lead to cognitive overload, reduced attention, fatigue, and ultimately a poor user experience. These concepts are further utilized in developing a section of the framework for human-centered robots.

**Perception**, or the process of understanding, is another subcomponent of human cognition. Some researchers discuss it as a combination of sensation and perception as separate entities, but generally, it refers to the entrance of information through human sensory channels (sensation) and the interpretation and assignment of meaning to that information by linking it to memory.

For instance, in interacting with an anthropomorphic robot, when feedback is received from its facial expressions, humans first intake the information through sensory channels (seeing it). Then, they must work to assign meaning to the observed information and comprehend it (perceive it). This second stage highlights the importance of interaction design, as poorly designed feedback can lead to misunderstandings, errors, and ultimately a poor user experience.

Perception, as a concept, is inherently complex and closely interlinked with other components such as attention, memory, and more. In earlier generations of technology interaction, researchers suggested presenting information in chunks and carefully selecting fonts and contrast to enhance comprehension. Following this path, contemporary focus includes improving the clarity of explanations provided by robots.

**Memory** refers to the process of storing and retrieving information. A well-functioning memory enables individuals to accurately interpret the meaning of sensory inputs. Researchers have explored various types of memory, including sensory memory, short-term memory, and long-term memory. In interactions with a robot, recognizing its face, recalling past interactions, and remembering the intent behind a specific movement are clear examples of how the concept of memory applies to human-technology interaction.

In such interactions, sensory memory, which temporarily stores sensory information for a few seconds, plays a critical role. It briefly retains sensory inputs provided by the robot, allowing essential information to be selected through the process of attention and subsequently processed for perception.

As a result, robots should deliver sensory feedback in a way that enables necessary information to be quickly transferred to working (short-term) memory when required. This means that the feedback must effectively engage the user's attention. Once the user focuses on specific information, it is transferred to short-term memory. The capacity of short-term memory is limited, traditionally estimated to hold around $7 \pm 2$ items/chunks.

The next stage is long-term memory, which is responsible for storing and later retrieving information over extended periods. For instance, in interactions with a humanoid robot, when we refer to leveraging past experiences for a new interaction, we are essentially discussing the retrieval of information from long-term memory. If specific information provided by the robot needs to be memorized, repetition or presenting it within meaningful interactive scenarios are effective methods for encoding it into long-term memory.

Overall, understanding the concept and processes of memory and designing interactions based on memory principles are crucial for enhancing the user experience in human-robot interactions. Interaction design that considers memory processes and the inherent limitations of human memory is one of the key factors in successfully human-centering robots.

Another subset of cognitive processes, both generally in individuals and specifically in interactions with robots, is **problem-solving and decision-making**. In this process, individuals actively think and utilize their cognitive resources to choose one option among those available and act upon it to influence their environment. For example, in an interaction with an educational robot, as previously discussed, when a human seeks to communicate with the robot or respond to its actions, the processes of problem-solving and decision-making are actively at play. Understanding the processes that lead to decision-making in humans can provide significant assistance in designing optimal interactions with robots.

In this context, other cognitive dimensions, such as attention level, mental fatigue, cognitive workload, and time pressure for making decisions, are also generally considered together. It is unlikely that decision-making and problem-solving in interactions with a robot occur in isolation. Therefore, human-centering interaction with a robot requires attention related to problem-solving and decision-making. For instance, the clarity of explanations provided about a robot's functionality is among the most critical principles to consider in human-centered design. This principle directly intersects with human processes of perception and decision-making.

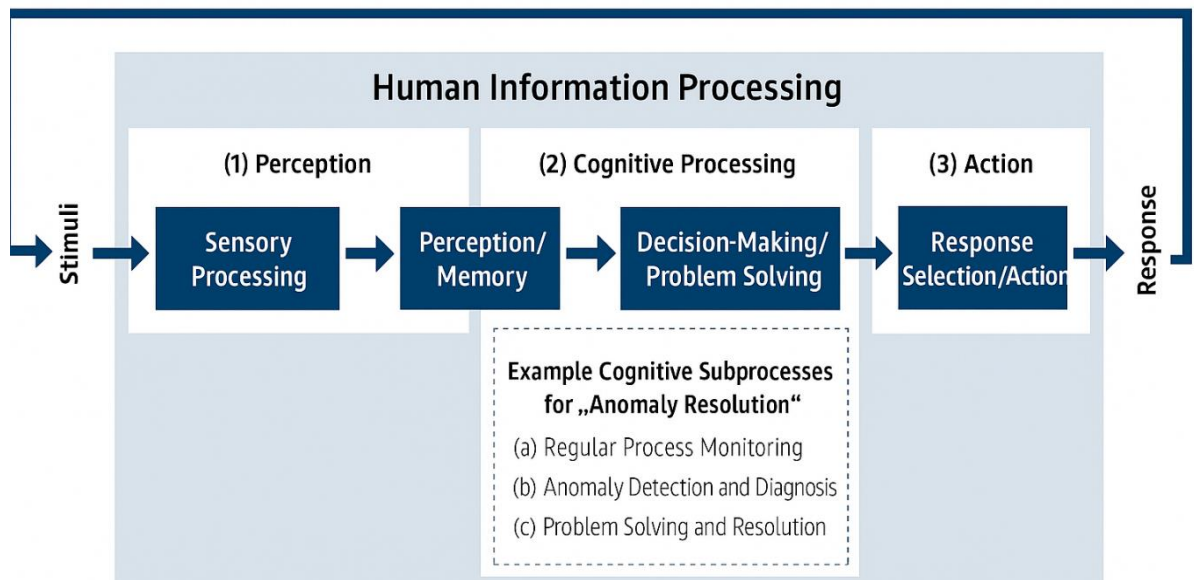

*Figure 4. Simplified version of Human information processing in interaction with an industrial robot. Adopted from (Morgenstern et al., 2024)*

To deepen the understanding of human cognition, researchers have utilized conceptual frameworks based on human cognitive principles in relation to technology. These models and frameworks aim to explain human behavior when interacting with technology and their performance during such interactions.

One of the most significant concepts humans use in interactions with robots is **mental models** (Andrews et al., 2023). In general, when a person encounters a robot for the first time or seeks an explanation for an unusual robot behavior, they heavily rely on mental models. Over time, as exposure and interaction with technology increase, these mental models—naturally dynamic in nature—develop and update. Therefore, designing robots in alignment with users' mental models can be a significant step toward providing a satisfactory user experience. Robots aligned with mental models enable smoother, simpler, and more user-centered interactions, reduce the cognitive load of interactions, help users better understand and learn the robot's functionality, and, on a broader scale, increase trust and acceptance of robots within society.

For instance, one of the most common mental model users have about robots is that they function like human assistants. In this case, to human-center this interaction, the robot should be able to understand natural speech and exhibit simple, predictable behaviors, much like an assistant.

When interacting with a robot through mental models, our conscious and unconscious cognitive processes are likely influenced by these models beforehand, guiding our behaviors during the interaction. Nowadays, researches show that many people might interact with robots based on incorrect mental models. For example, when a robot encounters an error during interaction, many users attempt to repeatedly pose commands to cancel the previous command. However, this approach is flawed because the robot, being frozen or unresponsive, cannot process any new commands, including the cancellation request.

Therefore, it is essential to design interfaces in a way that helps users form correct mental models of the robot or rectify their inaccurate ones. These interfaces aimed at correcting users' mental models could include simple, easy-to-follow instructions rather than requiring users to read lengthy text. Additionally, providing short videos demonstrating how the robot operates in specific scenarios has been identified as another effective method for fostering accurate mental models.

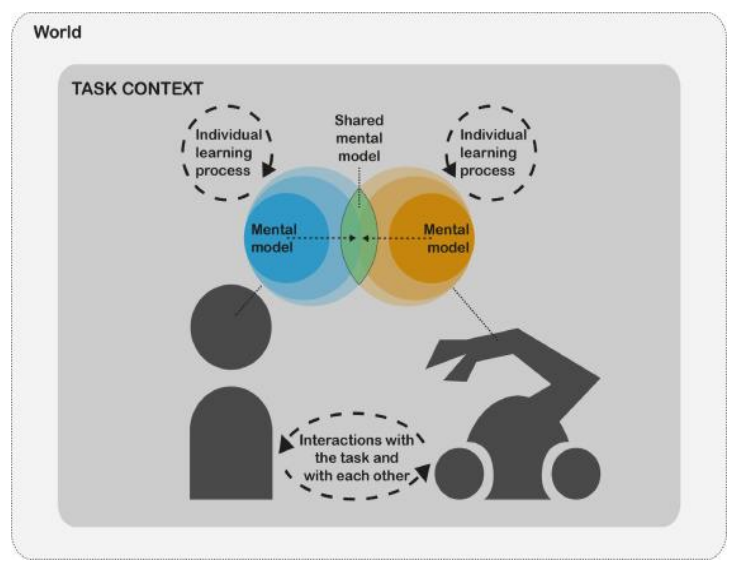

*Figure 5. Human mental model in HRI adopted from (Schoonderwoerd et al., 2022).*

Another cognitive framework for users interacting with technology (in this case, robots) is Donald Norman's theory of the two Gulf, also known as the **Gulf of evaluation and Gulf of execution theory** (Limerick et al.,

2014; Lo & Helander, 2005). The execution gap refers to the distance between the user's intention or goal and the action required to achieve it. The evaluation gap, on the other hand, refers to the distance between the system's actual performance and the user's perception of the outcome.

Human-robot interaction designers should aim to minimize these two gaps (gulfs). Achieving this can make human-robot interaction smoother and more effective. To reduce the execution gap, intuitive user interface designs, the use of standardized gestures and patterns aligned with user expectations, and simple implicit training are recommended. To address the evaluation gap, providing clear and prompt feedback after the robot's actions, ensuring the feedback aligns with user expectations and is easy to interpret, as well as displaying the robot's current status in plain and understandable language, are effective strategies.

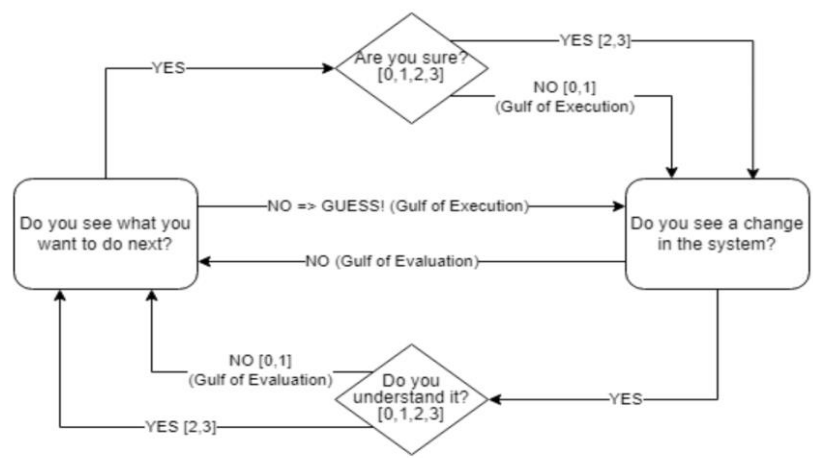

*Figure 6. The gulf of evaluation and execution. Adopted from (Ozsoy, 2023)*

There is another category of conceptual models and cognitive theories that can help human-robot interaction designers gain deeper insights into human behavior in specific contexts. These frameworks often focus on how the environment, tasks, and specific needs influence cognitive processes.

For instance, in **Situated Cognition Theory**, researchers believe that human cognition is heavily dependent on the environment and the context in which it occurs. Based on this well-established theory, efforts are underway to design and develop robots that can accurately recognize different environments and interact with users accordingly. For example, consider a language-teaching robot working with children to teach a second (non-native) language. If the robot understands that the child is currently in a restaurant, it could enhance effectiveness by teaching vocabulary and structures related to the restaurant and food. Similarly, imagine a medical robot that can assess the urgency expressed by a user (in this case, a surgeon) regarding the patient's condition and the criticality of a surgery. The robot could then adapt its behavior to prioritize faster actions to assist in an emergency surgery (Wang, 2024).

Another theory that can be used in human-robot interaction is **Distributed Cognition Theory**. According to this theory, cognitive activity does not occur solely within an individual's brain; rather, it emerges from the interaction of their brain with other individuals, tools, and the environment (Cavuoto & Bisantz, 2020).

For example, in a rescue operation where a robot collaborates with a rescuer, the robot scans the environment to find survivors and shares this information with the rescuer. The human rescue team then analyzes this data and, in coordination with the robot, advances the rescue mission. In this scenario, the entire cognitive process is distributed across multiple humans interacting with one another and with the robot. Understanding the concept of distributed cognition can guide efforts to find solutions that enhance the level of human-robot collaboration. Additionally, interfaces and interaction methods should be designed to ensure better management of information flow between humans and robots.

Two other cognitive theories, closely related conceptually and frequently referenced in human-robot interaction, are the **Extended Mind** and **Extended Cognition** theories (Drayson, 2010; Gallagher, 2013).

In the first, humans use robots as external resources to support their cognitive activities. That is, humans rely on robots to process information in certain scenarios. For instance, consider a situation where a student's academic progress is assessed by a teaching assistant robot, and only the final report is delivered to the human teacher, enabling him/her to make further decisions for the student. In this example, the aim is to empower teachers.

In the second theory, Extended Cognition, humans seek assistance from robots to expand their cognitive abilities. For example, when a teacher cannot independently evaluate students' performance, they enlist the help of a robot to effectively extend their cognitive capacity.

While it is challenging to draw a clear practical distinction between these two theories, the key difference lies in their emphasis: Extended Mind focuses on the expansion of the mind and cognition beyond the brain, while Extended Cognition emphasizes the use of external tools to process information. Like the previously mentioned theories and conceptual models, both of these frameworks can be effectively applied to human-centered robot design.

Another cognitive theory gaining attention in human-robot interaction, and which can be utilized to human-center such interactions, is **Embodied Cognition**. This theory derives its meaning from the human body and how it manages physical interactions with the environment (Lindblom & Alenljung, 2015). In this context, physical human-robot interaction is examined through the lens of this theory.

If an external object demonstrates how it operates within an environment, humans can more easily perceive its functionality and make decisions about interacting with it. For instance, in interaction with an educational robot, if the robot can clearly show its physical position relative to the language learner or if a robotic surgical assistant can indicate its next moves in response to the surgeon's actions, a seamless interaction is achieved. According to this theory, the way humans physically interact with a robot in a specific environment shapes their understanding of the robot's capabilities in performing tasks within that environment.

Another application of this theory in human-robot interaction lies in teaching a particular concept to an individual. For example, if a robot, through its movements in the environment, can elicit a physical response from the user without placing the entire cognitive burden of learning that movement solely on their mind, the learning process becomes easier. By engaging the individual's physical faculties in response to the robot's actions, the learning experience is facilitated. Consider a scenario where a robotic therapy assistant teaches a stroke survivor to regain movement in their limbs. In such a case, the robot's ability to guide physical movements effectively through its actions aligns directly with the principles of Embedded Cognition.

**2.2. Human Social considerations in HRI:**

After reviewing cognitive processes and related theories, it is essential to explore social processes concerning humans and technologies like robots. This approach enables us to understand concepts such as trust in robots and robot acceptance in society (Kanda & Ishiguro, 2017; Kim et al., 2018).

Humans are inherently social beings; they live together, work together, learn together, play together, and collectively form society. Technologies that have successfully fulfilled humans' social needs over the years have been widely accepted and utilized in communities. A clear example of this was during the COVID-19 pandemic, when social networks on smartphones became crucial in maintaining human connections during quarantine.

Understanding how humans naturally interact and identifying the social needs that robots can address to enhance social integration allows for better human-centered robot design. Several critical social attributes must be integrated into robots to ensure their acceptance in human societies; otherwise, these robots are likely to fail.

For instance, robots must support cooperation and social interaction among humans. Naturally, human presence in groups often revolves around collaboration and interaction, and any robot introduced into such settings must ensure constructive role (In the framework presented in the following section, this cooperation is considered essential for meeting certain higher-level human needs at a macro level). Additionally, it is crucial for robots to adhere to societal rules and norms. Robots should respect the social rules and standards that humans have developed and maintained, sometimes over centuries or millennia, rather than disrupting them.

Various sociological and social psychology theories have been developed to study human behavior in societies. Today, some of these theories can be leveraged in designing human-robot interactions. Designing a successful robot requires understanding these theories to create robots that align with and complement human behavior within societal contexts. By doing so, the initial steps toward human-centered robot design for societal integration and interaction with diverse individuals can be taken.

**Role theory** focuses on the roles individuals assume in social interactions. Humans naturally assign roles to others, including robots, and adjust their expectations based on these roles. This makes the consideration of a robot's potential role in society crucial in its design. For example, when designing an educational robot, it must be explicitly clear whether the robot is intended to act as a teacher or merely as a teaching assistant. Similarly, in therapeutic robots, clarity is needed on whether the robot will assume the role of a physician or serve only as a supporting tool for the physician.

In the context of human-centered interaction, ensuring role clarity for robots is a practical application of this theory. By transparently defining the robot's role in its design, we can help set realistic expectations, reduce confusion, and enhance the overall quality of human-robot interaction (Blaurock et al., 2022).

**Social Exchange Theory** posits that humans seek to maximize benefits in their social interactions, preferring relationships that are not only free from harm but also advantageous. Consequently, when designing robots, it is crucial to ensure that the interaction requires minimal effort from humans while offering significant rewards or benefits. If a robot fails to demonstrate its utility within a short period, its relationship with the user is unlikely to persist. This theory can justify certain elements of the framework presented in the next section, particularly the emphasis on the efficiency and effectiveness of robots for individuals, which forms the foundation of the framework. By aligning robot design with the principles of this theory, we can foster sustained and meaningful human-robot interactions (H. Kim et al., 2022).

**Expectation Theory** emphasizes that individuals have specific expectations about the behavior and responses of others, shaped by their prior experiences and social beliefs. For example, when we smile at someone we've just met, we anticipate a warm and positive response rather than a confrontational one. In line with this theory, the social behavior of robots should align with user expectations. For instance, if a user expects a robot to respond quickly (or conversely, slowly) to a command, the robot must meet that expectation precisely to ensure consistency and user satisfaction. Adhering to this principle in design enhances the likelihood of seamless and positive human-robot interactions (Kwon et al., 2016).

Since human-robot interaction differs somewhat from human-human interaction, as most humans encounter robots for the first time while they have had many prior human-human interactions, the concept of first impressions in a social interaction comes into play. One of the relevant theories in this area is the **Dual-process models of impression formation** (Koban & Banks, 2023). According to this theory, humans use two main pathways to form impressions of robots. The first pathway is a fast and automatic route. This route relies on automatic processing and emotions, typically used when individuals lack cognitive resources or time for deep analysis of the robot. The second pathway, in contrast, is a slow and analytical route that involves conscious and logical processing. This pathway is typically activated when the individual is motivated to interact or feels the need for a deeper performance analysis of the robot. In this case, the robot's performance and behavior are usually evaluated through the use of long-term memory.

In human-centered robot design, it is necessary to address both pathways, which have been well documented by social psychology theorists. That is, robots must create a positive first impression in the initial interaction, such as having an attractive appearance or performing human-like movements with a friendly face (provided these align with the expectations and role theories). This would address the initial impression phase.

However, for an analytical and deeper impression, clear performance and explanations for the robot's actions are required. For example, when robots are used in therapeutic or medical settings where long-term interactions are necessary, the appearance of the robot may no longer be as important as its clear performance for the user. On the other hand, if we design a robot that closely mimics human movements, it may not be necessary for it to resemble a human's face. As this discrepancy between performance and appearance could disrupt the first impression formed through the first pathway.

Another important theory in social sciences that can easily be applied in human-robot interaction is the **Social Comparison Theory**. According to this theory, users are likely to compare the robot's performance to that of a human or, if they have previous experience, to other robots. For instance, when explaining this theory in robot design, it is crucial to understand that robots which are meant to serve as assistants should not behave too similarly to humans. They must provide clear explanations about their supplementary and assisting role so that the user understands that the primary responsibility for performance lies with them, just as one would expect from an assistant in a social environment. In fact, this approach ensures that humans can recognize their own legal and social responsibilities as the primary individual in the interaction and understand that, in case of mistakes, they are fully accountable (Kamide et al., 2013).

Another important theory applicable in human-centered human-robot interaction is the **Social Norms Theory**. According to this theory, robots should clearly and transparently follow the cultural norms of the society and specific context in which they operate (S. Kim et al., 2022). For example, in cultures where eye contact is highly important for communication, a social robot should, first of all, have clear eyes and use them effectively in social interactions. In societies where maintaining a certain personal space is important when people are near each other, the robot does not need to get too close to an individual during interactions. The same example of cultural and social norms can be extended to a robot designed for a European culture, used in a Muslim country in Africa or Asia. Naturally, it is expected that the robot will respect at least the basic ethical and social norms of the Muslim community.

In the long-term use of robots, whether at home or in the workplace, according to the **Shared Intentionality Theory**, the robot must clearly engage in team tasks and its specific role in the required performance, demonstrating that it shares common goals and intentions with the user (Dominey & Warneken, 2011). It is through this alignment that a robot can be expected to be accepted in society. In applying this theory in the framework presented in the next section, we emphasize that the robot should collaborate in achieving high-level human needs, such as reducing poverty and hunger, and helping humans in their pursuit of the 17 United Nations Sustainable Development Goals (Haidegger et al., 2023).

### 2.3. The most documented cognitive and social constructs in HRI:

In this section, based on the cognitive and sociological concepts discussed in the previous two sections, we will review the most important conceptual structures that have garnered attention from researchers in human-robot interaction in recent years.

### 2.3.1. Personality:

Personality type refers to a set of relatively stable traits and behavioral patterns over time, reflecting how individuals think, feel, react to their environment, and engage in social interactions. Like other cognitive-social concepts, personality types have been studied and presented through various theories and models, with notable examples being the Five-Factor Model (Big Five) and the Myers-Briggs Type Indicator (MBTI) (Briggs, 1974; Soto & Jackson, 2013). According to a broad consensus among theorists, individuals with different personality types typically exhibit distinct patterns in information processing, decision-making, problem-solving, and behavior (Babamiri et al., 2024b). A well-known example is the difference between introverts and extroverts in interacting with their environment. These two groups often have varying emotional patterns and ways of regulating their feelings. In human-technology design, leveraging personality types can create a personalized experience tailored to each individual's personality, enhancing usability and satisfaction (Esterwood & Robert, 2021).

Research has shown that designing personalized user interfaces and experiences based on personality types can improve the efficiency and effectiveness of interactions. For instance, consider a robot interacting with an extroverted, socially enthusiastic person versus the same robot interacting with an introverted individual. If the robot quickly identifies these differences and successfully adapts its behavior, the interaction quality is likely to improve significantly.

On the flip side, we can also imagine assigning personality types to robots (Mou et al., 2020). Recall from the previous section that one of the main differences between robots and other AI-driven technologies is their physical embodiment. Now, envision a physically embodied robot with a distinct personality type. For example, how would an interaction between an introverted robot and an introverted human or between an extroverted robot and an introverted human play out? Can you imagine a future where robots with varying personality types interact with one another in a world where humans are no longer alone?

A 2018 study revealed that personality type in human-robot interaction (HRI) encompasses various dimensions. Approximately 41.5% of research in this domain focuses on human personality types, while 30.5% examines robot personality types and their impact on interactions. The remaining 28% explores similarities and differences in personality types between humans and robots and their effects on interaction (Robert, 2018).

Some personality-related outcomes studied in HRI include performance, robot acceptance, and trust in robots. The most frequently explored topic for both human and robot personality types is introversion versus extroversion. For example, in recreational and serious environments, extroverted robots are often perceived as more sociable and playful, aligning with the mental model typically associated with extroverted humans. An important consideration in robot design is assigning a personality type to the robot. One study found that in healthcare environments, individuals were more inclined toward extroverted robots, whereas in military settings, participants responded more positively to introverted robots.

Therefore, it is crucial to consider the interaction context and the personalities of individuals in the environment when designing a robot. This consideration becomes particularly significant in human-centered robot design. A designer must not only ensure the robot's effectiveness but also make an extra effort to ensure the robot is perceived appropriately by users and is compatible with their personality types and the interaction context.

### 2.3.2. Anthropomorphism:

Anthropomorphism, by a commonly agreed definition, refers to attributing human traits, characteristics, and behaviors to non-human objects, animals, or systems, such as robots and other technologies. It is often assumed that humanizing objects and technologies leads to better and more engaging interactions. However, some studies indicate that excessive anthropomorphism can have adverse effects on interaction. Moreover, distinctions have been made between varying levels of anthropomorphism, such as superficial versus deep anthropomorphism. superficial levels tend to have positive effects, while deeper levels may lead to negative outcomes. Understanding the features and theories related to anthropomorphism can significantly aid designers in creating human-centered robots (Złotowski et al., 2015).

Anthropomorphism is closely related to several social theories previously discussed. Based on an initial assumption rooted in the theory of mind, humans tend to actively ascribe mental states to non-human entities, especially when they expect those agents to perform complex behaviors typically expected from humans. Therefore, incorporating certain levels of anthropomorphism in designing robots that handle multiple tasks can be beneficial. However, determining the appropriate degree requires careful analysis of the relevant context, the individuals involved in the interaction, and the interaction environment (Atherton & Cross, 2018).

The uncanny valley effect suggests that if a robot resembles a human too closely but fails to behave naturally like human—consider the difference between robotic limb movements and the smooth, natural motions of human limbs—it can cause discomfort for humans (Vaitonytė et al., 2023). A similar concept can be found in the cognitive dissonance hypothesis. If a robot's appearance or behavior resembles that of a human, but its functionality significantly deviates, it may result in cognitive dissonance.

In the next section, where the chapter's main framework is presented, it is highlighted that robots must align with human cognitive and social theories. For instance, if a robot is anthropomorphized for close human interaction, the social theories discussed here must guide its design.

In general, a significant portion of past studies has focused on anthropomorphism in the physical appearance of robots. However, there have also been studies that examined behavioral anthropomorphism. For instance, features such as the ability to make ethical decisions, deceive humans, engage in human-like conversations, or demonstrate accountability for their actions are among the behavioral characteristics in robots. When present, these features make robots more human-like.

### 2.3.3. Trust:

If we were to identify a single key concept at the center of human-robot interaction, many researchers would agree that it is human trust in robots (Hancock et al., 2011). Unlike trust between humans, which often involves emotional connections and complex feelings, trust in robots predominantly relies on assessing their technical capabilities and the quality of their interactions. In essence, the technological aspect outweighs the emotional one (Law & Scheutz, 2021).

Trust also has an initial component, referred to as "initial trust," which generally forms before any interaction with the robot. Various psychological, social, and individual factors influence this stage. Trust can be regarded as one of the primary factors driving the acceptance of intelligent robots in society.

In the framework presented in the next section, implementing human-centered principles and meeting user needs—whether at the individual level or on a broader scale involving multiple human-robot interactions within society—is directly influenced by users' trust in robots. Consequently, a specific position has been allocated for trust in the framework's outer layer.

### 2.3.4. Acceptance:

The acceptance of a technology, such as robots, in society encompasses a broader perspective of human-robot interaction, including factors like culture and the socio-economic conditions of the community, as highlighted by researchers. In general, acceptance refers to users' willingness to use robots and integrate them into their daily lives and various environments. Clearly, this concept is a subsequent step after trust in robots (Esterwood et al., 2022).

In addition to the macro-level cultural and sociological factors mentioned above, trust in robots during interactions also significantly influences their acceptance. In fact, a person's decision to embrace robots as collaborators, assistants, or companions is a multifactorial parameter. This decision depends not only on the robot's technical and functional capabilities but also on various cultural, psychological, and sociological factors.

Several theories address technology acceptance in general and robot acceptance in particular, the most notable being the Technology Acceptance Model (TAM) and the Unified Theory of Acceptance and Use of Technology (UTAUT) (Ahmad, 2015; Davis, 1989). These models emphasize different factors contributing to acceptance. For instance, according to TAM, perceived usefulness and perceived ease of use are crucial. This model suggests that if a robot provides tangible benefits and is easy to operate, its acceptance likelihood increases (Y. He et al., 2022).

UTAUT, considered a more advanced version of TAM, also incorporates social and motivational factors. Building on these foundational models, newer models have been continually developed. Examples include the Persuasive Robot Acceptance Model and the Human-Robot Collaboration Acceptance Model. Specific acceptance models have also been proposed for robots in specialized contexts, such as the Service Robot Acceptance Model (Chen et al., 2024), Robot Acceptance Model for Care (Felding et al., 2023), and Service Robot Acceptance in Museums (Wong & Wong, 2024).

Clearly, acceptance is a broad and overarching concept. In the conceptual model presented for human-centered robot design in the following section, acceptance is positioned at the top right as the ultimate outcome of integrating robots into society and their acceptance. Its predictive parameters, such as efficiency, effectiveness, and ease of use—well-documented in the aforementioned models—are placed in the lower sections of the pyramid within this chapter's conceptual model.

## 3. Human-Centred AI Framework with Focus on HRI

### 3.1. Introductory points: From Usability/UX to Human-centred AI

In the first section, we discussed the growth process of robots and introduced social robots as the offspring of the Industrial Revolution. Following this, we provided a brief overview of their most significant applications in close interaction with humans, such as in healthcare systems, educational institutions, and other similar domains. In the second section, we explored general cognitive and psychological mechanisms of humans, both broadly and specifically in their interactions with robots. In this section, the focus towards presenting an integrated framework for human-centred robot design.

Approximately 60 years ago, when robots were first being designed, the fundamental principles of their design naturally focused on functionality. This meant that if a robot could successfully carry out the specific task intended by its designers, it was considered a well-designed robot. Over time, this purely function-driven perspective, gradually evolved (Boy, 2017).

In the evolution from industrial robots to interactive robots, and eventually to today's social robots powered by large language models, various related disciplines—such as psychology, industrial design, human factors, and others—also experienced significant maturation. The integration of these disciplines with the traditional principle of functionality highlighted the increasing importance of humans as both users and collaborators with these robots. Around 40 years ago, it became evident that the mere ability of a robot to perform its intended function was no longer sufficient; the experience of the human interacting with the robot also began to take centre point. One of the earliest and most significant concepts in this context, which had already emerged in the related field of human-computer interaction a few years earlier, was usability. This concept dates back to the 1970s and with efforts from the International Organization for Standardization (ISO) in the 1980s to standardize its definitions and application, usability gained broader recognition. According to the ISO standard, formally introduced in 1998, usability encompasses three key subscales: effectiveness, efficiency, and user satisfaction. This marked a pivotal moment, as it emphasized that not only the effectiveness and efficiency of a system were critical, but also the satisfaction of the user had to be considered in system design. Following the introduction of the usability concept and its design counterpart, user-centred design (UCD), in 1986 by Donald Norman and colleagues, researchers began to advocate for the evolution of UCD. They argued that UCD, which until then had drawn its strength from the core principles of usability, should be expanded to human-centred design (HCD) to better address the broader context of human needs and experiences (Cooley, 2000; Ferwerda et al., 2024).

Overall, the prevailing belief was that this definition went beyond usability, encompassing a broader range of human needs in the design process—both in general product design and, more specifically, in the design of technological products. However, both usability and human-centred design (HCD) fundamentally centered around the idea of considering humans, their needs, and their limitations during the design process.

In the late 1990s and early 2000s, researchers began to argue that usability, as a metric for HCD, was no longer sufficient on its own to fully capture human needs in design—particularly in the context of technology. Users, also have needs for pleasure, aesthetics, and beauty. What they meant was that, instead of solely focusing on usability, greater attention should be given to the entire process of the user's perceived experience when interacting with technology and robots. This led to the emergence of the concept of user experience (UX). One of the key pioneers of this conceptual evolution was Donald Norman, who asserted that designing a product according to usability principles could make it functional yet unattractive. This marked the beginning of critiques against purely usability-driven design and the development of the newer concepts of user experience and emotional design. During the same period, other related concepts, such as hedonic design, Kansei engineering, and affective computing, were also introduced. While these concepts had slight differences, their shared focus on addressing the aesthetic, emotional, and psychological needs of technology and robot users far outweighed their distinctions. From the 1970s onward, the concept of usability gradually evolved toward user experience and emotional design, with increasing applications in the design of robots—specifically, in the design of human-robot interaction. Over the past several decades, as human-robot interaction has evolved from usability and user-centred design (UCD) to user experience (UX)—and now, in the era of AI-powered robots gaining widespread adoption, to human-centred AI (HCAI) design of robots—a single overarching principle has remained constant: prioritizing humans and their needs in the design of robots (Sayago, 2024; Siricharoien, 2024).

But what exactly does "need" mean? What human needs are robots expected to address? With this introduction, we delve into the concept of human-centred AI design for robots, its various dimensions, and the frameworks researchers have proposed to guide this process.

**3.2. Selected Framework for the HCAI Topic: Learns generally, Applies specifically to HRI:**

This section aims to review several models, guidelines, and frameworks proposed in the broader field of Human-centered AI (HCAI), which, while not all explicitly focused on human-robot interaction, offer valuable insights. In the end, the key parameters identified by various researchers and organizations as essential criteria for HCAI will be consolidated. It is evident that these models and frameworks share a high degree of similarity, and their collective outcomes will serve as the foundation for developing a new framework tailored specifically to human-robot interaction in the subsequent section.

**3.2.1. AI4 People Framework**

Luciano Floridi and colleagues (Floridi et al., 2018), in their framework described as an ethical framework, summarized insights from reviewing various documents from organizations and individuals into five core components. These components include beneficence (do good), non-maleficence (do not harm), human autonomy, justice, and explainability. The first component focuses on the benefits of AI-based systems, highlighting aspects such as promoting well-being, preserving dignity, and sustaining the planet. The second component emphasizes the importance of preventing harm, with discussions on topics like privacy, security, and capability caution. The third component addresses the significance of human autonomy in delegating decisions to AI or retaining decision-making authority. Users must have the option to override AI decisions if they wish to do so. In the fourth component, the framework underscores AI's role in fostering fair distribution of resources, ensuring shared benefits, and promoting common well-being across all individuals globally. Finally, the fifth component identifies explainability as a complement to the four preceding principles—beneficence, non-maleficence, autonomy, and justice. This section also elaborates on explainability's key sub-elements, including accountability, intelligibility, and transparency.

**3.2.2. Three-Layer Framework by Wei Xu et al.**

In 2019, this researcher introduced his conceptual framework for the first time, consisting of three core components: User, Technology, and Ethics (Xu, 2019). This framework outlined seven primary objectives for human-centred AI systems: Usable AI, Scalable AI, Trustworthy AI, Empowering User AI, Useful AI, and Controllable AI. A closer examination of the details within these seven objectives reveals key elements such as privacy, fairness, explainability, controllability, trustworthy, accountability, comprehensibility, unexpected behaviour bias, keeping humans in the loop, and ease of learning. Similar to other frameworks, these elements collectively contribute to and define the overarching seven objectives of this framework.

**3.2.3. Microsoft Guideline**

Saleema Amershi and her colleagues introduced a 18-point guideline developed at Microsoft for Human-AI Interaction (Amershi et al., 2019). While this guideline was not explicitly crafted for human-centred AI systems, a review of these 18 principles reveals that they align closely with the goals of human-centred AI design. These principles are categorized into four stages: Before interaction (initial), Throughout the interaction (during interaction), At the time of the error occurrence (when wrong), and in Long-term use (over-time). Key recommendations from this guideline during interaction with AI systems include ensuring that these systems adhere to the user's social and cultural norms while also avoiding biases and promoting fairness. Equally notable are the suggestions for handling errors in AI systems. These guidelines emphasize the importance of addressing errors thoughtfully, as such situations are critical for fostering effective human-robot interaction. This aspect, in particular, has not received the same level of attention in other research as it does here, highlighting its significance.

**3.2.4. Google People + AI Guidebook**

This guideline, first published by Google in 2019 and last updated in 2021, serves as a guide for integrating AI into products (Google, 2021; Yildirim et al., 2023). The handbook is structured into six main chapters that cover topics such as user needs, mental models, explainability and trust, feedback, user control over the system, and error management. A review of these six categories reveals principles such as user augmentation, responsible AI, fairness, social benefit, safety, accountability, privacy, sustainability, human wellness, and reliable AI. These principles align closely with those addressed in other guidelines and frameworks within the field.

**3.2.5. A Systematic Approach to Human-Centred AI Proposed by Greece Researchers**

George Margetis and his colleagues (Margetis et al., 2021), including Professor Constantine Stephanidis from Greece, have identified 6 fundamental concepts for human-centred AI by reviewing several guidelines and frameworks:

Explainable AI and Human-in-the-Loop, Semantic Cognitive and Perceptual Computing, Visual Predictive Analytics, Interactive Machine Learning, Federated Learning, and UX Design for AI.

In their detailed explanation of the above concepts, they have made an effort to keep the core HCD (Human-Centred Design) model as the central framework. One of the key points in this guideline, like the primary model of HCD, is the attention to context and the specific needs of users in that particular context. In the next step, they emphasize ergonomic needs and the user experience alongside functionality and then focus on solution design and solution evaluation with the active participation of all stakeholders, which is another important aspect of their guideline. In the six-fold classification presented, concepts such as fairness, explainability, transparency, trust, beneficence, human rights, human control, moral and ethical considerations, following social values, well-being and empowerment, safety, and others are effectively highlighted.

### 3.2.6. IBM: Design for AI Activity Guidebook

In 2019, this guidebook was published by the IBM group (IBM, 2019). The researchers from this group believed that before designing AI-based systems for humans, it is essential to first understand humans and their needs. A significant portion of this guidebook is dedicated to the concept of ethics in the design of these systems, with the main categories including accountability, value alignment, explainability, fairness, and user rights. In the first section, the guide addresses the company's overarching policies regarding various levels of accountability for AI products. Unlike other guidelines that focus on designing interactions based on accountability, this guide takes a broader perspective on the topic. The second section, however, shifts focus to the products themselves. The authors of this guidebook believe that products should align with the social norms and values of the consumer group. In the third section, the concept of explainability is discussed, with a focus on the ease of understanding, recognizing, and interpreting the decision-making processes of AI systems. In the fairness section, it is noted that humans are susceptible to biases, and AI systems must be designed to avoid discrimination between different user groups. In the final section, like many other guidelines and frameworks, the guide discusses privacy, data protection, and user control.

### 3.2.7. Ben Shneiderman's Framework for Human-Centred AI

The focus of this guide is a shift from one-dimensional thinking to a two-dimensional approach, where high levels of automation can coexist with the necessary human controls in the system (Shneiderman, 2022). According to this framework, AI systems should not simply mimic humans but should be designed and implemented to empower them. This guideline, along with its supplements, is elaborated in a book by the author and addresses several important concepts related to empowering humans. These include, among others, reliability, bias, transparency, fairness, explainability, understandability by providing new UI, safety specially at a macro level in the form of safety culture, trust, accountability, no deception, etc. As mentioned earlier, the focus of this researcher is on shifting from a one-dimensional perspective (which solely emphasizes increasing the level of automation) to a two-dimensional approach (which integrates enhancing human autonomy alongside automation). This is because the researcher believes that the former perspective is inaccurate, incomplete, and misleading for human-AI interaction. True empowerment of humans can only be achieved when, alongside the increase in automation, individuals also retain control over it, so that qualities like intelligence, creativity, self-confidence, and others are preserved and strengthened. Otherwise, with a purely quantitative increase in automation, humanity will be diminished.

**3.3.8. HCAI Essential Approach by Focusing on 6 Grand challenges**

In an international collaboration (Ozmen Garibay et al., 2023), 26 renowned researchers in the fields of artificial intelligence, human factors, and human-centred design identified six main challenges in human-centring AI, which include the following:

1. Focus on Human Well-Being
2. Responsible Design
3. Privacy
4. Design Based on HCD (Human-centred Design)
5. Appropriate Governance and Oversight
6. Respecting Human Cognitive Capacities

In the process of introducing these principles for human-centring AI systems, these individuals considered an AI to be human-centred if it possessed at least characteristics such as ethics, fairness, and an understanding of human conditions. Furthermore, they mentioned in part of this article that AI should be able to assist humans in achieving the United Nations' development goals that humans are striving to reach. Examples of these goals include poverty reduction, increased justice, better healthcare and education systems, and more. In interpreting the six main challenges mentioned above for focusing on enhancing human well-being, aspects such as harm avoidance, trust, accountability, agency, UX and effect, and multiple optimizations are highlighted. For an AI to be responsible, four key features are considered essential: ethics, accountability and liability, explainability and transparency, and fairness. Additionally, regarding the topic of privacy, the authors have pointed out that the primary challenge lies with data, specifically the data related to humans that are stored within AI systems. Six different types of privacy are addressed in this paper including the right to be left alone, the right to limit access to the self, the right to secrecy, the right to control over personal information, the right to protect person hood, the right to intimacy.

For the fourth challenge, which is about following HCD principles for AI systems, the initial framework of the Double Diamond model is discussed. According to this model, which has been widely examined in human-centred design, the entire AI lifecycle—from problem definition, data collection for model training, model execution, and validation, to the application of the model—should involve active participation from all stakeholders. Regarding the fifth challenge, which focuses on governance and higher-level regulatory concepts, the authors emphasize that governance is not solely about rigid legislation. Instead, they approach it tentatively, viewing it as a dynamic and evolving process. In the sixth challenge, which introduces design based on human cognitive abilities, the emphasis is placed on human-AI collaboration across various aspects of life, such as employment. It highlights the importance of task allocation between humans and AI, establishing a balance between empowering individuals and the potential loss of their professional abilities, the replacement of humans by AI in various jobs, and other related concerns.

## 3.3. Human-Centred AI in the Literature of HRI

It can be concluded from the previous section that the researchers have identified over 20 different parameters as criteria for human-centred AI-based products. However, there are not many articles that specifically focus on presenting these parameters as frameworks or conceptual models for human-robot interaction. In this section, we will review articles in this field. Hongmei He and colleagues have explored the challenges and opportunities of human-centred AI for trustworthy robots and autonomous systems (TRAS) (H. He et al., 2022). They believe that a robot should possess five key characteristics: safety, security, fault tolerance and health check, easy to use and effective HMI, and compliant with law and ethical expectation. At the end of this paper, a new acceptance model is presented based on these five parameters.

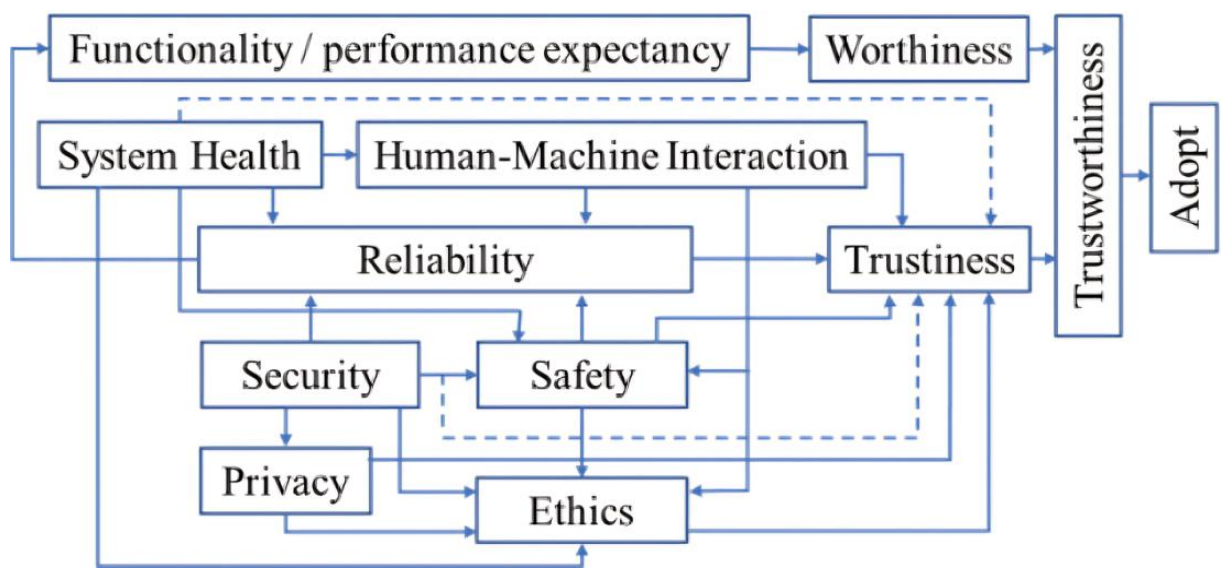

*Figure 7. Acceptance model for trustworthiness of robots and autonomous systems. Adopted from (H. He et al., 2022)*

In another paper, recently published as an invitation to a half-day workshop as part of the side programs of the MUM ZOZ4 conference, general points regarding the human-centred design of social robots were highlighted (Y. Zhang et al., 2024). A deeper examination of these points was conducted through a review of the papers presented at the conference. These points broadly included efficiency, effectiveness, meaningfulness, ethical considerations, intuitiveness, empathic, responsible AI, enhance the quality of the life, socially appropriateness, trustworthiness, transparency, safety, privacy and security, usability, acceptability, personalized and adaptive interaction, which were identified by the workshop's organizing team, comprised of researchers from various European countries, as critical aspects of human-centred design for social robots. In a critical paper by Stephen Doncieux and colleagues published in 2022, the authors explored the differences between robots and other AI systems (Doncieux et al., 2022). One of the key distinctions highlighted was the physical embodiment of robots in close proximity to humans, which had been discussed earlier in the same chapter. However, following the presentation of these fundamental differences, the paper reviewed the main considerations for making robots human centred. According to this research team, one of the essential factors is the robot's ability to learn—specifically, learning from its mistakes and correcting them. Additionally, they emphasized that robots should be versatile and adaptive. Another key requirement identified was termed "understanding humans", which involves recognizing and responding to human behaviours effectively. Furthermore, the paper underscored the importance of robots making their movements comprehensible to humans, ensuring clarity and predictability in their actions. Performing physical movements, especially in scenarios where physical interaction with humans is crucial, was highlighted as another key parameter for human-centred robot design. The ability to convey the reasoning behind movements and actions through interfaces was also emphasized as a critical factor. In this context, concepts such as explainable interfaces were indirectly addressed, ensuring that users can understand the robot's behaviours and decisions.

The final section of these requirements was dedicated to ethical considerations, which included topics such as building trust in robots and ensuring their acceptance by users. To elaborate on ethical principles, the authors made extensive reference to existing ethical guidelines—such as those published by the High-Level Expert Group on AI (HLEG-AI) (AI, 2019). For instance, the HLEG-AI guidelines include elements such as fairness, transparency, privacy, safety and reliability, human agency, beneficiary for human and environment, accountability.

In another article by Anastasia and colleagues, the presence of social robots and their impact on society were analysed through the lens of existing ethical recommendations and guidelines, including the Five Senses Ethical Framework (Ostrowski et al., 2022). While reviewing key ethical considerations highlighted in previous studies, the authors sought to propose a new guideline called the HRI Equitable Design Framework, aimed at designing human-centred robots. The new guideline was developed to address areas that had received less attention in prior research. These critical aspects included autonomy and control, transparency, and deception, providing a more comprehensive approach to ethical and human-centred robot design.

### 3.4. A Proposed Framework Based on the Literature

At the end of Section 3.1, the question was posed: **What are the human needs in interacting with a robot?** What specific requirements does a human have when engaging with a robot or a group of robots that should be considered in the process of human-centred robot design?

To address this question, Sections 3.2 and 3.3 presented key concepts from the domains of human-centred artificial intelligence and human-centred design for AI-powered robots. These sections aimed to provide foundational insights for understanding the human-robot interaction context. In this section, an effort has been made to consolidate the information available in the relevant literature and propose a **new and comprehensive framework** for **human-centred AI robotic**.

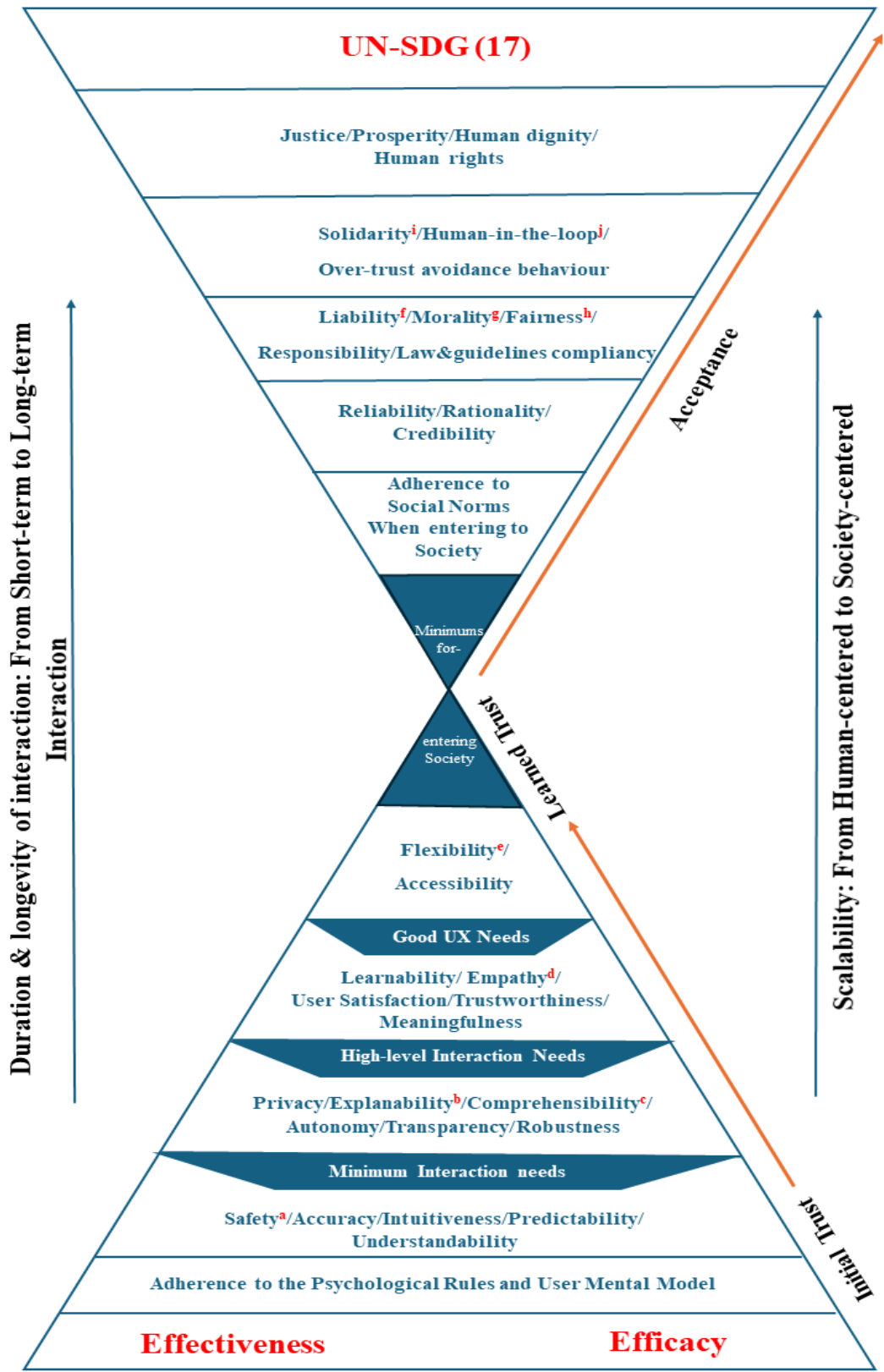

*Figure 8. A new proposed framework for Human needs in Human-centered AI robots*

a:Assurance; b:Explicibility; c:Interpretability; d:Emotional; e:Adaptability; f:Accountability; g:Ethical behaviour; h:Inclusiveness and non-biased behaviour; i:Coherency; J:Human-controllability

The proposed framework, named the **Dual Pyramid for Human-centred AI robotics**, begins at the lowest point of the first pyramid, representing the basic human needs in a one-on-one, short-term interaction between a human and a robot. As guiding by arrows on the right and left, the longer the interaction or the more robots and humans involved (i.e., the robot's integration into society), the more comprehensive and enriched the list of needs that interaction designers must consider becomes. The connection point between the two pyramids is conceptualized as the readiness for the robot's entry into society (interactions with larger human populations). In general, the **lower pyramid** focuses more on human-robot interaction needs at the individual level, while the **upper pyramid** addresses human needs related to the robot's presence in society. Each of these pyramids is composed of **six layers**, and it is recommended that designers move sequentially from the lowest layer of the first pyramid to the highest layer of the second pyramid. Additionally, the **context** for which the robot is being designed is crucial in determining the relevance of each layer. For example, for a multipurpose social robot, all—or at least many—of the 12 layers may need to be considered. However, for a task-specific industrial robot or a robotic assistant in an autonomous vehicle, some layers may hold less significance. Within each layer, the importance of certain parameters may vary depending on the context. For instance, the concept of safety in the case of a social robot designed for an operating room differs significantly from the safety requirements of a stationary social robot like [Furhat](#).

At the lowest layer of the bottom pyramid, the minimum design requirements for a robot include **effectiveness** and **efficiency**. These two elements can be considered the most fundamental aspects, which justified the presence of the earliest robots several decades ago by meeting these basic needs. In other words, if a robot is intended to enter the market as a product, and it fails to meet the minimum human needs of **effectiveness** and **efficiency** in the specific context for which it was purchased, it cannot even meet the first step of being human-centred. Following this foundational layer, based on our team's analysis, the second layer highlights the importance of adhering to certain **cognitive principles** and **aligning with the user's mental models.** This means the robot should match the user's mental expectations of what a robot should be and adhere to basic cognitive principles, such as avoiding excessive cognitive load or unnecessary stress on the individual. After these first two layers, the third layer focuses on a set of key **interaction parameters** between the robot and a human. Depending on the robot's design purpose and its specific context, some or all of these parameters may hold varying levels of importance or relevance. The fourth layer introduces **higher-level interactive needs**, where the significance of each parameter, as in the previous layer, may differ depending on the context. The fifth layer addresses concepts related to **UX** of a robot. Finally, in the sixth layer of the bottom pyramid, the robot's ability to effectively communicate with a wide range of individuals is emphasized. This layer, in connection with the subsequent layer, lays the groundwork for the robot's presence in broader contexts and **enter to the society.**

This means that for a robot to operate on a large scale within society, it must first fulfil certain concepts from the lower layers, ultimately culminating in the ability to interact effectively with a diverse range of users (flexibility). The foundation of the upper pyramid begins with **basic social needs** and adherence to certain **human social norms** and progresses toward **collaborating with humans on a societal scale (multiple human-multiple robot interaction)** to achieve the **United Nations' Sustainable Development Goals (SDGs)**. In other words, the first step for a robot, in the initial layer, is to demonstrate **efficiency and effectiveness** within a specific context for individual users. The final step in human-centring robots involves their ability to **collaborate with humans at a**

**macro level** to contribute to larger societal objectives. For example, the collective presence and widespread interactions with robots are expected to help significantly reduce **poverty (SDG 1)**, facilitate access to healthcare services **(SDG 3)**. The most tangible examples of human-centred robots contributing to broader societal goals include environmental sustainability **(SDGs 11 and 13)**, reducing inequalities **(SDG 10)**, fostering peace **(SDG 16)**, promoting quality work and innovation in industry **(SDGs 8 and 9)**, and enhancing access to education **(SDG 4)**.

**3.5. A Glimpse on Human-Centred Criteria in human-robot interaction:**

In this section, an attempt has been made to briefly review the concepts presented in the framework. The definitions provided are aimed at aligning with the context of human-robot interaction, meaning that the same terms may have slightly different definitions in interactions between humans and other technologies.

**Effectiveness:** The robot's ability to perform tasks assigned by humans, meaning the extent to which the robot can accomplish the task expected to do.

**Efficiency:** The optimal performance of the robot with the minimum frequency (repetition) of receiving command and minimal energy consumption (lower rate of error for repeating the same command).

**Adherence to Psychological Norms and Mental Models:** The extent to which the robot adheres to general principles, cognitive abilities, and limitations of humans and aligning with their mental models. For example, the robot should not impose excessive cognitive load on the human. Additionally, due to the $7\pm2$ limitation in short-term memory, providing simultaneous information may cause problems for the human.

**Safety and Assurance:** Ensuring physical safety and preventing harm or injury to the user.

**Intuitiveness:** Ease of using the robot without the need for complex instructions or training, usually related to the robot's user interface for interaction.

**Accuracy:** The robot's accuracy in performing tasks and providing error-free responses to user commands.

**Understandability:** The simplicity and clarity of the robot's actions or behaviour that allow the user to have an overall understanding (not necessarily specialized or deep) of it.

**Predictability:** The robot's ability to act in such a way that its subsequent actions and movements can be easily predicted.

**Explainability/Explicability:** The framework similarly considers them interchangeability, and in general, it refers to the robot's ability to provide simple explanations to the end user (non-specialized).

**Comprehensibility/Interpretability:** Although both concepts, which are considered synonymous here, are similar to the two points above, the subtle difference lies in the robot's ability to provide more technical and in-depth explanations to interested individuals or specialists.

**Transparency:** The ability of the robot to communicate the reasons behind its actions to the end user. Clearly, the distinction from the previous points lies in the focus on providing information rather than necessarily ensuring understanding it. Therefore, it might not be sufficient on its own.

**Consistency:** The robot's ability to maintain its performance quality at an acceptable level over a significant period of time.

**Privacy:** The robot's ability to ensure the preservation of data collected from the user.

**Good Repair Strategy:** The robot's ability to compensate for errors or failures in its performance and restore the user's trust.

**Trustworthiness:** In human-robot interaction, it refers to the user's confidence in the robot's capabilities and behaviours. In the framework, it is a higher level of the previously introduced parameters.

**Meaningfulness:** The robot's ability to provide value to the user and perform goal-directed behaviours related to their needs, such as making a significant difference in the overall quality of the user's life.

**User Satisfaction:** The robot's ability to satisfy the user in one or more specific contexts.

**Empathy/Emotional:** The robot's ability to establish an emotional connection with the user, including understanding their needs, emotional states, and responding with empathy, similar to human interactions.

**Robustness:** The robot's ability to maintain consistent performance in an environment characterized by uncertainty and unexpected conditions.

**Flexibility/Adaptability:** The robot's ability to engage with a wider range of people or address diverse needs from an individual.

**Accessibility:** The robot's ability to meet the specific needs of individuals requiring assistance.

**Adherence to Social Norms in Human Societies:** The robot's ability to follow socially accepted norms across different cultures and human societies.

**Reliability:** The robot's ability to deliver consistent performance to all users at a societal level.

**Rationality:** The robot's ability to show wise behaviour in society, such that a significant number of individuals attribute reasoning, thoughtfulness, and intelligence to it.

**Credibility:** The robot's ability to act wisely in society such that a significant number of people trust its actions and behaviour. It's a one-more step from the last one.

**Accountability/Liability:** The robots' responsibility regarding their behaviours in society. This means that both the robot and its designer must be accountable for the robot's actions and its presence in the community.

**Morality/Ethical:** Adhering to minimum ethical requirements in different societies. Imagine a robot from a European country is produced for the market in a Muslim country. Can it behave in the same way in a Muslim country at the societal level as it does in Europe?

**Fairness/Inclusivity/Non-Biased:** The robot's ability to treat all users equally, without discrimination based on gender, race, skin colour, or other factors.

**Laws and Guidelines Compliancy:** Robots must adhere to higher-level laws and national and international guidelines for participation in society.

**Responsibility:** Developed within the concept of Responsible AI, refers to the robot's commitment to acting in alignment with the broader needs of society.

**Over-trust Avoidance:** Robots must act in a manner that prevents excessive trust or over-reliance from society. This does not mean intentional malfunctioning, but rather reminds users of the importance of human involvement in the relationship.

**Human-in-the-Loop and human controllability:** Refers to the inclusion of humans in the decision-making process, with the human being as the final decision maker.

**Solidarity/Coherence:** The important aspect of robots as companions, collaborators, and contributors to human endeavours aligned with societal goals.

**Justice:** The robot's belief in human values such as justice and functioning on a large scale to help humans achieve it.

**Prosperity:** The robot's ability to assist humans in achieving prosperity and societal well-being.

**Dignity:** The presence of robots in society should contribute to preserving and enhancing human dignity within the community.

**Human Rights and Rules:** Robots should not hinder human progress toward essential human rights, such as democracy.

**UN-SDGs (17):** Robots should help society achieve the broader goals of the United Nations Sustainable Development Goals (SDGs), such as reducing inequality, providing adequate healthcare, ensuring quality education for all, promoting safe transport, and creating a sustainable society.

## 4. Human-Centred HRI in the Field

So far, it has been clearly emphasized that robots must be more than mere technology, their integration into human environments must encompass not only optimal user experiences but also a deep focus on addressing human needs through the concept of human-centred AI design. In this section, an overview is provided of the role of robots as collaborators and companions in specific environments. This review seeks to critically analyse the strengths and weaknesses of previous studies, examining them through the lens of the parameters outlined in the earlier framework.

### 4.1. Human-Centred AI robotics in Healthcare Systems

The first part of this chapter also highlighted that social robots hold significant potential to assist healthcare systems in areas such as rehabilitation, autism, diabetes, cancer, dementia, and more. In essence, these robots can help alleviate the heavy workload and address the shortage of human resources in healthcare systems. Additionally, they can serve as companions, monitors, and aides to patients. If you recall the framework presented earlier, effectiveness and efficiency were identified as the minimum necessary criteria for human-centred robot design. While there is extensive literature on the effectiveness and efficiency of deploying robots in healthcare systems, the results remain mixed. Although some studies reported positive outcomes, others failed to demonstrate adequate effectiveness or efficiency. To evaluate effectiveness, it is essential to consider the initial goal of deploying the robot in a healthcare setting and assess whether the presence of the robot achieved that intended objective or not. In 2022 a review article examining the use of social robots to improve the well-being of elderly individuals highlighted findings from 12 studies on effectiveness and 10 studies on efficiency (Mahmoudi Asl et al., 2022). The results of this review revealed both ends of the spectrum—demonstrating both effective and not effective outcomes. At least seven studies reported improvements in emotional states and mood. Additionally, three studies focused on engagement, six on social interaction, one on increased job satisfaction among healthcare staff, and one on reduced self-reported pain levels among elderly individuals—all of which indicated varying levels of effectiveness. On the other hand, at least four studies examining quality of life (QoL) and four studies focusing on reducing depression reported a lack of effectiveness.

This pattern of effectiveness and efficiency across various outcomes follows a relatively similar trend. Another review article examining the use of robots in therapeutic interventions for children reported more favourable results regarding effectiveness (Triantafyllidis et al., 2023). Out of 13 included studies, 11 demonstrated positive effectiveness, while two reported a lack of effectiveness.

Overall, the integration of robots into healthcare systems has generally accounted for the highest levels of human needs within the framework of human-centred robotics. For instance, few fields can rival healthcare robotics in terms of their alignment with global development goals, such as the United Nations' Sustainable Development Goals (SDGs). A paper published in October 2024 introduced a philosophical discussion on the design of healthcare technologies and robots, emphasizing patient needs and preferences (Gogoshin, 2024). The study concluded that, since patients typically enter healthcare systems seeking treatment, only those preferences (and needs) that directly contribute to resolving their medical issues should be prioritized in the design of such technologies. It argued that while considering all patient preferences might seem more ethical on the surface, it is not necessarily required if the primary focus is on addressing their health concerns effectively.

Thus, between the two levels of human needs—minimum (effectiveness and efficiency) and maximum (interaction with robots to achieve the United Nations' Sustainable Development Goals)—there are numerous factors that have been considered as human-centred parameters in previous literature. These include privacy, accountability, trustworthiness, fairness, transparency, explainability, human-in-the-loop, liability, responsible AI, and safety.

It seems that the healthcare system is the industry where the most efforts have been made to human-centred robots. The likely reason for this, as opposed to military environments, is the diversity of individuals involved, ranging from children to the elderly, with varying levels of health. However, the primary reason can be attributed to the high demand within the healthcare system for robots to support its workforce. Few industries, aside from workplaces, have research focused on the necessity of robots within them, but the healthcare industry stands out in this regard. A review of various studies on the presence of robots in healthcare leads to the conclusion that, for several reasons, including the high workload of personnel, the aging population, and more, the presence of robots in this sector is essential. These two factors—the longstanding and essential presence of robots on one hand, and the wide range of individuals with different characteristics they interact with on the other—have led to more effective steps being taken towards human-centred AI robots in this industry compared to other sectors.

**4.2. Human-Centred AI robotics in Military**

Similar to the previous case, when discussing the application of robots in military environments, we must start from the lowest point of the framework, namely effectiveness and efficiency, and then progress upwards through the other elements of the framework.

Chohan and his colleagues have addressed the deployment of robots and their effectiveness in enhancing military capabilities, believing that the use of robots is a forward-looking solution for military organizations (Chohan et al., 2023). However, it seems more appropriate to consider the effectiveness and efficiency of these military robots in a specific task or under particular conditions when discussing the minimum criteria for human-centeredness. It would be more beneficial not to generalize about their overall effectiveness. It would also be better to introduce and use quantitative indicators for measuring the effectiveness of military robots. For instance, how much error is acceptable? How much cognitive resource utilization from the user is reasonable? After the minimum requirements related to effectiveness and efficiency presented in the framework, the most significant topic concerning the human-centered AI of military robots pertains to the ethical concepts in their usage. Within this subset, similar concepts such as moral acceptability, military ethics, and responsibility have also been discussed and examined by the authors.

For instance, regarding ethical responsibilities, there is a mix of beliefs where arming robots is considered both ethical and unethical depending on the perspective (Taylor, 2021). However, in the case of semi-autonomous robots controlled remotely, the consensus is that the ethical responsibility for their actions lies with the military commanders who operate them from outside the battlefield. Some also believe that the manufacturers of these robots bear a degree of responsibility as well. On the other hand, with fully autonomous armed robots, the issue of accountability becomes significantly more complex. Some argue that this level of uncontrolled automation should not exist at all, as it raises profound ethical concerns about an autonomous armed robot fighting against humans. However, some ethicists believe that the presence of fully autonomous robots might eliminate the need for human involvement on the battlefield altogether, potentially ending wars without human bloodshed. Additionally, beyond ethical considerations, the vulnerability of robots and military facilities to electronic warfare threats, as well as broader issues of privacy and cybersecurity, have been identified as critical challenges in human-centred design for military robots (Borges & Rosado, 2024). Scholars often reference the vulnerabilities observed during the ongoing Russia-Ukraine war to underscore the importance of addressing these concerns. Another emerging application involves robots designed to protect humans as bodyguards. According to the parameters of the proposed framework, the primary concerns related to such protective robots revolve around ensuring their safety mechanisms and preventing them from inadvertently harming other individuals while safeguarding their designated human subjects (Duarte et al., 2022). In this context, ethical considerations for such robots are inherently tied to safety. In another section of the literature on human-centred robot design in military contexts, some researchers argue that the prospect of fully autonomous robots operating without human oversight is highly unlikely. They suggest focusing instead on fostering effective collaboration and interaction between humans and robots. Within this perspective, transparency in military robots emerges as a critical yet underexplored parameter (Lakhmani et al., 2020). This concept encompasses related aspects such as understandability.

In essence, when a robot is expected to collaborate with military personnel, it must provide clear and comprehensible explanations regarding its reasoning and decision-making processes. The rationale for emphasizing transparency in military robots is closely tied to supporting human operators in maintaining situational awareness and ensuring that humans remain in the loop.

### 4.3. Automated Vehicles

In reviewing the literature on the presence of robots in automated vehicles, two primary perspectives regarding their role emerge. The first group views automated vehicles as inherently robotic systems, arguing that the vehicles themselves can be classified as a type of robot (Mosaferchi et al., 2024). The second group, which has garnered greater support among researchers, posits that the presence of robots within automated vehicles serves to make them more human-centric and socially integrated (Large et al., 2019). The second perspective emphasizes that while automated vehicles may not inherently a robot, the inclusion of robots within them plays a pivotal role in enhancing their social and human-oriented functionality. Key reasons for incorporating robots into automated vehicles include fostering trust among passengers, assisting with navigation, serving as an explainable interface between humans and the vehicle, and managing emergency situations in a more human-like manner. This perspective highlights the role of robots as facilitators of social interaction and user support within automated vehicles.

Preferring the second category does not necessarily mean that the first perspective is incorrect. Self-driving cars can indeed be seen as robots, as they, like robots, process data from their environment (through sensors and scanners), use AI, and then take action accordingly. However, this section of the book views robots in self-driving cars as interface between humans and vehicles to facilitate their socialization. The difference between these perspectives can lead to variations in the concepts presented in the framework of this chapter. For example, discussing the efficiency of automated cars as robots differs from considering the efficiency of robots as interface between passengers and vehicles, as the parameters and tasks defining their success or failure are different.

Among the key parameters of the framework highlighted in the second look are transparency, understandability, explainability, privacy, accountability, and responsibility. It could be argued that since a vehicle is a mobile technology, the most important cognitive parameter related to humans in connection with vehicles can be reasonably approximated as the driver's situational awareness (Lu et al., 2017). In integrating this cognitive concept with principles related to human-centred AI robot inside AVs, Chen et al. have proposed a situational awareness model, which underscores the heightened significance of transparency and explainability compared to other parameters of the framework, particularly within the specific context of automated vehicles (Chen & Barnes, 2023). Key reasons for incorporating robots into automated vehicles include fostering trust among passengers, assisting with navigation, serving as an explainable interface between humans and the vehicle, and managing emergency situations in a more human-like manner. This perspective highlights the role of robots as facilitators of social interaction and user support within autonomous vehicles. Although the importance of parameters such as safety should not be overlooked, there is a need for clarification. While safety may arguably be the most critical issue regarding autonomous vehicles, its significance, when considering the role of robots as socializing agents of autonomous vehicles—essentially their role in making autonomous vehicles human-centred—may be less pronounced than parameters like transparency and explainability. Of course, these concepts are inherently interconnected and cannot be entirely separated. Given the importance of transparency and explainability in autonomous vehicles, often delivered to humans via humanized interfaces, the authors have attempted to propose a new conceptual model for the acceptance of autonomous vehicles.

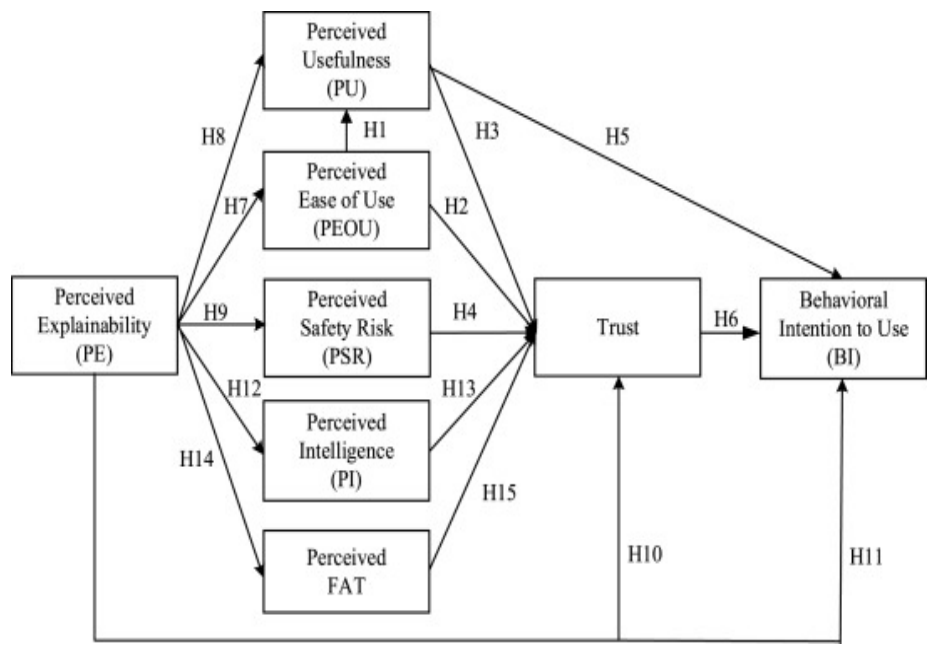

*Figure 9. New AV acceptance model enriched with perceived explainability. adopted from (T. Zhang et al., 2024)*

However, if autonomous vehicles are considered purely as autonomous robots, a different interpretation of the framework may be necessary. Nonetheless, parameters such as efficiency and effectiveness in driving, safety in accident prevention, and, at higher levels, collaboration with humans to achieve United Nations development goals seem more significant. Additionally, beyond the importance of explainable AI (XAI), some researchers have addressed autonomous vehicles from ethical perspectives in accidents and concepts like accountability and others (Schäffner, 2024).

**4.4. Toy and Entertainment Robots**

Entertainment robots, a significant subset of social robots, are often studied with a focus on their technical aspects. However, within the framework of human-centred AI robot design, there are key considerations specific to these robots. The prevailing belief in this context is that effectiveness—measured by the robots' ability to entertain children, adolescents, and even specific groups of adults—should be considered the primary parameter for human-centred AI design. On the other hand, a different group of researchers argues that since the primary objective of these robots is to entertain children while unconsciously learning an educational lessons, the most critical human-centred design parameters lie beyond effectiveness. They contend that in this case emotional connections with these robots is of far greater importance (Hung, 2020).

Clearly, unlike contexts where explainability is a key focus, as mentioned earlier, the emphasis here shifts significantly due to the unique nature of the application. As highlighted above, some researchers refer to these robots as serious toys aimed at fostering meaningful play, with goals such as identifying cognitive or behavioural challenges in children and older adults and enhancing their abilities (Winfield et al., 2022). In this regard, meaningful play aligns with the broader concept of meaningful interaction, which is vital for the human-centred design of robotic toys (Ihamäki & Heljakka, 2024). The more these robots seek to collect user data—particularly from users with specific limitations or vulnerabilities—the greater the ethical concerns surrounding privacy, as emphasized in the proposed framework.

Alan F. T. Winfield and colleagues, in 2022, developed a framework for assessing the ethical risks of entertainment robots in the context of responsible robotics (Winfield et al., 2022). This framework categorized the ethical risks associated with entertainment robots, such as robot pets, into four major areas:

1. Physical Risks, such as battery overheating.
2. Psychological Risks, including addiction to the toy or children being deceived by the robot.
3. Privacy and Security Risks, such as misuse or leakage of personal information.
4. Environmental Risks, including the lack of repairability or recyclability of the robots.

This framework underscores the need for careful consideration of these risks to ensure responsible and ethical deployment of entertainment robots as the name of the framework suggests, psychological issues arising from the use of entertainment robots have received significant attention. For instance, the risk of over-trust, a parameter previously identified in this chapter's framework, has been categorized as a high-severity ethical risk. While most of the parameters outlined in the framework are considered to pose moderate risks, two other factors—security against hacking attacks and low transparency—have been highlighted as highly critical. Although it is reasonable to assume that the risks associated with toy robots might vary across the four categories of the framework depending on the specific design and use case, it is generally acceptable to consider this framework a valuable tool for assessing the risks—primarily ethical—of toy robots. Additionally, privacy concerns related to entertainment robots have also been a focus of researchers in the field of human-centred robotics design in other studies.

**4.5. Industrial Collaborative Robots**

Recall from the early sections of this chapter that the first generation of robots introduced in the automotive industry marked the beginning of the human-robot interaction concept. Undoubtedly, these robots were designed to overcome the limitations of human physical strength and fatigue, while enhancing productivity on assembly lines (Babamiri et al., 2021). From the perspective of the proposed framework, the focus of this early generation of robots predominantly revolved around efficacy and efficiency, which are foundational elements of the framework. At most, these robots could be linked to concepts of safety, but as time progressed and newer generations of robots emerged, additional concepts relevant to human-centred robotics have come into play. For instance, the integration of robots into industries today and their seamless collaboration with humans should ideally contribute to the United Nations' Sustainable Development Goals (SDGs) (Guenat et al., 2022). Specifically, Goals 8 (Decent Work and Economic Growth) and Goal 9 (Industry, Innovation, and Infrastructure) are directly related to the presence of these robots, and even Goal 1 (No Poverty) can be indirectly tied to their impact on economic productivity and societal well-being. Today, a new generation of industrial robots has emerged, moving beyond robotic arms confined to safety fences, working in isolation. These new robots operate in close proximity to human workers, collaborating with them directly. This new category of robots, referred to as collaborative robots (cobots), has fundamentally shifted the paradigm of human-robot interaction in workplaces.

As we move from robotic arms toward "Computers/Robots as Social Actors", the increasing proximity and interaction between robots and workers necessitate the consideration of a broader range of parameters in the human-centred robotics framework. For example, the privacy implications of interacting with these next-generation robots in workplace settings were explored by Henning and colleagues in 2024. Their research highlights the evolving challenges of privacy in the context of these collaborative robots. Efforts to make industrial robots more human-centred have also contributed to the emergence of the concept of Industry 5.0, as a progression beyond Industry 4.0. This paradigm shift signifies a move away from solely prioritizing productivity and profitability. Instead, it places a heightened emphasis on the role of humans, ensuring that they remain at the loop, fostering a more inclusive, ethical, and sustainable industrial ecosystem.

In 2024, Liang C.J and colleagues (Liang et al., 2024), reviewed the ethical challenges associated with the presence of AI-augmented robots in workplace environments and identified nine key challenges in this domain, including "Job loss, Data privacy, Data security, Data transparency, Decision-making conflict, Acceptance and Trust, Fear of surveillance reliability and safety, and Liability". These elements have also been incorporated into the various classifications within the framework presented in this chapter. As the likelihood of communicative collaboration increases, the transparency of robots is proportionally enhanced. In a 2022 review article, E. Coronado and colleagues expanded on the concept of human-robot interaction quality in industrial settings, focusing on human-centred factors relevant to robots in the context of the Fifth Industrial Revolution (Coronado et al., 2022). Key factors for human-centric design in these robots include safety, trust, acceptance, adherence to human cognitive principles such as mental workload, attention, mental models, and situational awareness, along with several UX- and hedonomic-related concepts such as emotional experience and affect.

Overall, in this framework, which is specifically designed for an industrial environment based on Industry 5.0, efficiency and effectiveness are introduced as fundamental indicators for a human-centred robot. Christian Levesque and his colleagues have introduced a concept called Better Work in work environments influenced by human-centred technologies (Lévesque et al.). The core of this concept is focused on eliminating the social and economic risks faced by workers and, in fact, empowering them, which aligns with the highest levels presented in the framework of this chapter. An associated framework, known as Enhanced Human-Automation Symbiosis (EHAS), Framework (Yaqot et al., 2024), focuses on topics such as society-centred technologies in the Fifth Industrial Revolution, human rights, and digital ethics—areas that are well-supported in the framework presented in this chapter. A comprehensive review of the presence of collaborative robots in industry convinces us that, although efficiency, effectiveness, and safety can still be considered the primary parameters for human-centred AI robotics, as the level of collaboration and interaction increases, concepts such as transparency, explainability, understandability, and others also become crucial. Additionally, at higher levels, the collaboration of robots in achieving UN Sustainable Development Goals and similar objectives plays an important role.

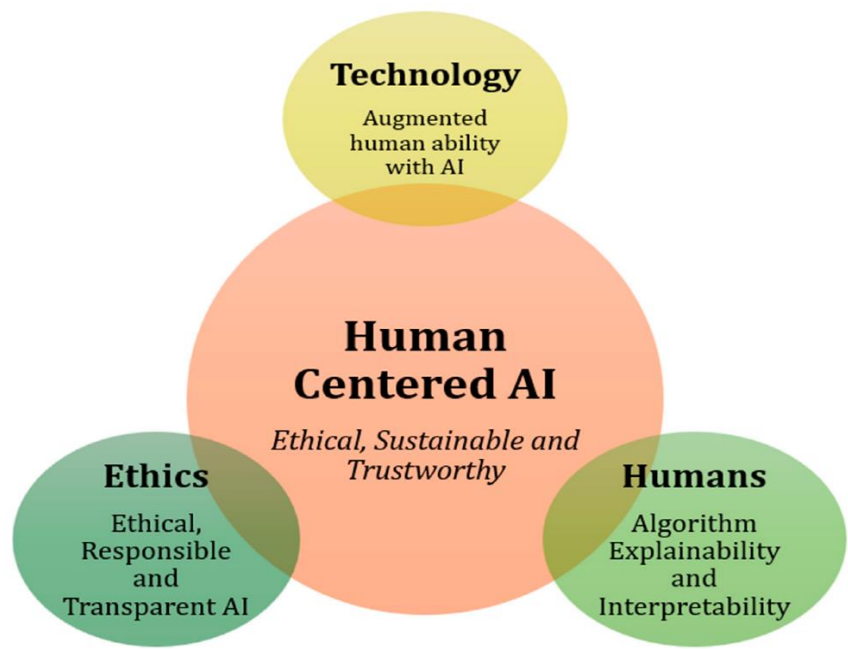

*Figure 10. Human-centered AI in Industry 5.0. Adopted from (Martini et al., 2024)*

## 5. Future Trends in HCAI in HRI

### 5.1. HCAI and HRI in the era of Generative AI (LLMs)

Up to this point, the key aspects of human-centric robotics aimed at optimizing human-robot interactions have been thoroughly analyzed. However, with the advent of a new generation of generative AI technologies, the landscape has grown increasingly complex. The remarkable advancements in large language models and their expanding range of applications have ushered human-robot interaction into a new era (Dimitropoulos et al., 2023). This development can be viewed as an advantage, but it also brings inherent limitations and challenges that warrant attention. For instance, among the benefits is the expansion of the potential use cases for robots into areas that were previously inaccessible due to constraints in natural language processing capabilities. For instance, robots equipped with large language models (LLMs) can now engage in more natural conversations with customers or even patients. Previously, voice interactions often fell short of resembling human-like communication. These LLM-powered robots, thanks to memory capabilities in models like GPT-4, can manage longer and more extensive sessions with users. This can open the door to more personalized applications, such as tailored education for children or therapeutic sessions with patients, further enhancing their effectiveness in advancing human-centred design and interaction (Onorati et al., 2023).

Social robots equipped with large language models (LLMs) possess near-human levels of social intelligence due to the extensive training of their language models on massive datasets of human-generated text. This endows them with the ability to engage in more human-like interactions. They can better understand users' emotions, participate in multi-layered conversations, and even provide contextually appropriate responses tailored to individuals' emotional states. On the other hand, unlike language models that are trained exclusively on text and thus have limited capabilities, multimodal models—capable of simultaneously processing text, images, and audio—have enabled robots to engage in richer, multidimensional interactions.

The ability to better detect emotions with multimodal models, such as vision-language models like CLIP, will certainly enhance interaction. In the near future, robots might even be able to communicate with individuals who are mute, solely through understanding their unique body language. However, alongside all these advantages, we must not overlook the new challenges and limitations that arise in the pursuit of human-centred robotics with the presence of this type of generative AI. Certain inherent drawbacks of language models, such as hallucinations, will likely be introduced to robots as well. If these occur repeatedly, they could erode user trust and confidence. Additionally, the biases and injustices inherent in large language models, which have been discussed extensively, could also pose a significant challenge to the human-centred approach to robotics. Another challenge is the significant processing time required by these robots, which is primarily due to the heavy computational demands of language models. This can lead to delays and make interactions with humans feel unnatural, ultimately disrupting the user experience. In conclusion, while the integration of LLMs and multimodal models has added new capabilities to robots, it is crucial to also address the challenges they bring—challenges that were rarely encountered in traditional robots—on the path to making these robots more human-centred.

## 5.2. Embodied Intelligence

If you recall from the early sections of this chapter, one of the key differences between human-robot interaction and human interaction with other technologies was linked to a concept called embodiment. It was noted that what sets robots apart from other technologies is their physical presence and the ability to interact with humans through their physical bodies. In connection with this concept, there is also the theory of embodied cognition and another framework known as embedded intelligence (Roy et al., 2021). These theories emphasize that human cognition does not occur solely within the brain; instead, the external environment and the way the body interacts with that environment have a significant influence on intelligence and cognitive processes. Building on this theory, researchers are working on a new generation of robots that can effectively learn from environmental experiences. Through this type of environmental learning, these robots can respond adeptly to human needs in complex environments. Such robots can be viewed as a significant step toward human-centred robotics by enhancing user experience and fostering greater trust and reliability.

Robots, through the integration of intelligence within their physical embodiment, strive to learn from their environment and respond to human needs more fluidly and effectively (Long et al., 2023). This new generation of robots no longer relies solely on pre-learned data and traditional machine learning techniques but instead can readily acquire new knowledge through reinforcement learning.

This learning can even occur via the simulation of human movements. For instance, by observing a human, understanding their actions through multimodal models, and attempting to replicate those movements, the robot can assess whether the action is considered "learned" based on receiving a reward for correctly executing it. Researchers believe that this type of robot, through collective interaction with other robots, holds the potential to integrate into society, following the theory of collective intelligence. Such robots can teach and learn from one another, further expanding their capabilities. Referring back to the framework discussed in the previous section, this new generation of robots is expected to embody concepts such as solidarity with humans, liability, rationality, and the overarching principle of not removing humans from societal participation (human-in-the-loop). Consequently, the parameters for human-centring these robots are considerably more complex than those of earlier intelligent robot generations. However, alongside these advancements in robotics and embedded AI, the challenges and concerns related to human-centring also grow. These range from foundational elements of the framework, such as safety and privacy, to higher-level considerations that were rarely addressed with previous generations of robots but are now increasingly pertinent.

## 6. Conclusion

The primary aim of this chapter was to explore the concept of human-centred AI in robotics. Specifically, it sought to highlight the importance of designing AI-powered robots with a focus on human-centric principles. To achieve this, the chapter provided an introduction to the history of robotics and their wide-ranging applications while addressing the core human needs in interactions with robots. Section three of this chapter introduced a new framework addressing a comprehensive list of human needs in interactions with robots, both at the individual level (micro) and within broader societal contexts (macro). According to this framework, the minimum requirement for human needs is the robot's effectiveness and efficiency in performing a specific task. On the other end of the spectrum, the maximum requirement involves solidarity between robots and humans to achieve the United Nations' Sustainable Development Goals (SDGs). Naturally, the framework also accounts for intermediate needs between these two extremes, such as safety, privacy, explainability, and other essential factors. The chapter concludes with an overview of relevant laws, regulations, and guidelines, as well as a forward-looking perspective on the future of human-robot interaction. The key highlighted in this chapter is the significant influence of context on the selection of human-centred AI parameters. For instance, the concept of human-centred design for robots in the healthcare system is likely to differ significantly from that applied in the automotive or military industries. So, in the future adopting this framework for each context (e.g., human needs in each context) is necessary. As an initial step, our team is working to validate this framework for using HCAI robotic in educational context (specific learning disorder in student) by a qualitative study with specific end-users.